# Virtual Reality Wireless Mobile Walkthrough Framework

**By**

Ghada Mohamed Fathy

**Under The Supervision of**

Prof. Dr. Reem Bahgat

Prof. Dr. Walaa Sheta

A Thesis Submitted to the

Faculty of Computers and Information

Cairo University

In partial fulfillment of the Degree of

Master in Computer Science

Faculty of Computers and Information

Cairo University

Egypt

**2014**

# IN THE NAME OF GOD, MOST GRACIOUS, MOST MERCIFUL…

# CERTIFICATE

I certify that this work has not been accepted in substance for any academic degree and is not currently being submitted in candidature for any other degree.

Any portions of this thesis for which I am indebted to other sources are mentioned and explicit references are given.

**Student Name:** Ghada Mohamed Fathy

**Signature:**

# ACKNOWLEDGEMENT

*First and foremost, I deeply thank Allah for his virtues and graces on me.*

*Secondly, I would like to thank my family for their infinite support and help; I dedicate all my achievements to them.*

*My sincere thanks to my supervisors from which I learned a lot of things..... Great thanks to Prof. Reem Bahgat for her valuable advices, kind guidance and cooperation through the work.*

*Special thanks to my great professor Prof. Walaa sheta for his cooperation and motivation during my research work and thesis writing. I really enjoyed working with him.*

*I would like also to thank Dr.Hanan; not only for the knowledge I acquire from her to enhance my research, but also for her guidance to enhance the way I work and serving my career.*

*I want to express my deepest gratitude to my friends especially to Zeinab for her helping and encouragement.*

# ABSTRACT

The past few years have witnessed a dramatic growth in the number and variety of graphics intensive mobile applications, which allow users to interact and navigate through large scenes such as historical sites, museums and virtual cities. These applications support many clients and impose a heavy requirement on network resources and computational resources. One key issue in the design of cost efficient mobile walkthrough applications is the data transmission between servers and mobile client devices. In this thesis, we propose an effective progressive mesh transmission framework that stores and divide scene objects into different resolutions. In this approach, each mobile device progressively receives and processes only the object's details matching its display resolution which improves the overall system's response time and the user's perception. A fine grained cache mechanism is used to keep the most frequently requested objects' details in the device memory and consequently reduce the network traffic. Experiments, in simulated and real world environment, are used to illustrate the effectiveness of the proposed framework under various settings of the virtual scene and mobile device configuration. Experimental results show that the proposed framework can improve the walkthrough system performance in mobile devices, with a relatively small overhead.

# LIST OF PUBLICATIONS

G.M.Fathy, H.M.Hassan ,W.M.Sheta, R.Bahgat, "Efficient framework for mobile walkthrough application," *Pervasive and Mobile Computing*, Sep. 2014

# TABLE OF CONTENTS

# LIST OF TABLES

# LIST OF FIGURES

# LIST OF ACRONYMS

| | |
|---|---|
| BMS | Broadband Mobile System |
| CC | Cloud Computing |
| CPM | Compressed Progressive Mesh |
| CPU | Computing Processing Unit |
| CSMA/CA | Carrier sense multiple access with collision |
| CU | Connectivity Updating |
| CUG | Collaborative Virtual Environment |
| DAOA | Degree of Attention about Object's Attributes |
| FIFO | First In – First Out |
| 4 G | Fourth Generation Network |
| GU | Geometry updating |
| IBR | Image Base Rendering |
| iOS | iphone operating system |
| KPCA | K. Mean + principal Component analysis |
| LOD | Level Of Details |
| LRU | Least Recently Used |
| MAR | Mobile Augmented Reality |
| MLM | Most Liklihood Movement |
| MPCM | Modified Compressed Progressive Mesh |
| NFU | Not Frequently Used |
| NRU | Not Recently Used |
| PCR | Progressive Connectivity Representation |
| PGR | Progressive Geometry Representation |
| PM | Progressive Mesh |
| RAND | Random |
| RTP | Round Trip Time |
| TCP/IP | Transmission Control Protocol / Internet Protocol |
| 3D | Three Dimensions |
| VR | Virtual Reality |

# CHAPTER 1

# INTRODUCTION

This chapter gives an introduction about the proposed framework, lists our motivations, and design challenges. In addition, it provides a reader's guide.

# 1  Introduction

## 1.1 Overview

The recent proliferation of mobile phones has created a unique opportunity for researchers to use all their capabilities to provide new applications. Streaming and sharing different kinds of content such as pictures, documents, objects and videos have become more popular than ever. Immersing multiple users in a 3D world, in which each user simply uses a smart phone, has become possible with the evolution of today's technologies in terms of mobile computing and computer networks. Although we have witnessed these advances, mobile devices still lack the proper resources to run graphic intensive applications, such as walkthrough systems. In general, a walkthrough application is a shared virtual environment where users at their devices interact with each other over a network, each user is symbolized by an entity called avatar. Users normally navigate through the scene, perform various actions and interact with the other users within the same area of interest. The virtual environment must be rendered in real time to display a consistent view for all users. As a result, excessive resources will be required to avoid performance bottlenecks and maintain a reasonable quality of service, typically in terms of response time, latency and rendering speed (number of frames per second).

To overcome these limitations, this thesis presents a framework that reduces the amount of transmitted object meshes through the wireless environment; and also matches the object data with the resolution of the graphics hardware. This allows the system to provide a proper rendering speed and acceptable quality of service. The proposed framework consists of three main modules: First, progressive meshes module which divide objects mesh into several partitions and transmit only the appropriate resolution (partition) according to the distance between the user's view point and the visible objects. Here, the system determines the user's moving direction and transmits continuously from a low, base mesh, to high resolution [1]. Second, the caching mechanism allows a client to utilize its memory and local storage to cache current visible objects that are likely to be visible in the near future [2]. Our cache replacement policy is based on the assumption that a farthest object from the user will not be requested again in the near future. Hence, when a free cache space is needed, the outermost object is selected. Third, constrained object's resolution selection module is used to identify the highest resolution of the object's mesh that will be transmitted, based on the device capabilities such as the screen size and network bandwidth. Thus, it avoids the transmission and processing of unnecessary data.

## 1.1 Motivations

The growing usage of mobile devices, power phones and smart phones in virtual reality application brings the need for building efficient applications for this domain. This environment has a set of challenges which makes it interesting area for working on. The challenges from the working environment are:

- Wireless network's limitations (connection instability, and limited bandwidth).
- Device limitations; mobile devices have limited resources (memory, processing, and power). Even with the emergence of new high specification devices, they are still not in wide use.
- Application data, the size of the virtual world is quite large as we have scenes that contain a huge number of 3D objects. Thus, the virtual reality environment contains a big amount of data that need to be saved.

## 1.2 The Proposed Solution

Our proposed solution is based on integrating progressive mesh, caching and constraint object resolution techniques to develop an efficient framework that accommodates the phone's limitations (processing, power consumption, and memory).

To build our solution we used three main modules: the client module, the wireless medium (Uplink, downlink) module and the server management module as shown in Figure 1.

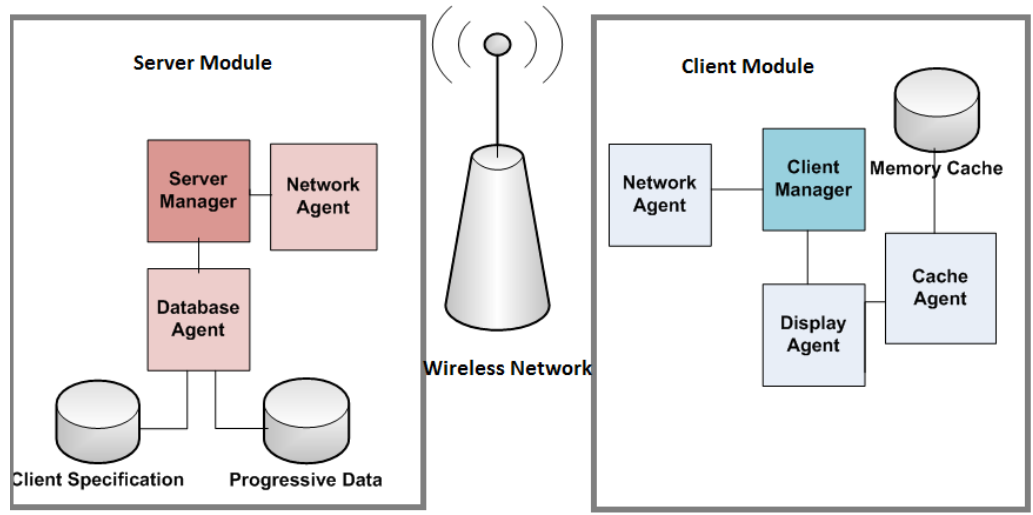

Figure 1: The framework components

**The client module**

The client module simulates the mobile device side and controls navigation, rendering, communication, and cache memory. The client module consists of four agents:

- *The client manager agent* it coordinates all other agents at the client side. It is responsible for simulating user navigations and preparing objects resolutions requests to be sent to the server by the communication agent.

- *The communication agent* it handles all communications with the server via the wireless medium. It maintains separate queues for sent and received requests that are processed by the cache agent to store/retrieve objects data.

- *The Cache agent* it stores the data received from the server and allocates free space according to a certain replacement policy. It also searches the cache memory for a certain object resolution and en-queues a sever request to the communication agent's sending queue in case of not finding that object.

- *The Display agent* it renders the scene objects on the device's display.

**Wireless Medium Module**

CSMA/CA (Carrier Sense Multiple Access/Collision Avoidance) is a protocol for carrier transmission in 802.11 networks. Unlike CSMA/CD (Carrier Sense Multiple Access/Collision Detect) which deals with transmissions after a collision has occurred, CSMA/CA acts to prevent collisions before they happen.

CSMA/CA protocol is simulated to handle data transmission between Server and client module [**3**].

**Server Management Module**

The server management module runs on the server side. It consists of three agents:

- *The server manager agent* coordinates all other agents at the server side.

- *The database agent* manages storage and retrieval of the scene objects' progressive records as well as the client's mobile specification.

- *The communication agent* handles all communications between the server and the clients through wireless medium. Two queues are maintained to handle both sent and received requests. Progressive data records are transmitted via a dedicated link to each client, while clients' requests are queued in the receive queue.

## 1.3 Thesis Organization

The thesis is organized as the following:

• **Chapter 1 Introduction:** Gives an introduction about the proposed framework, lists our motivations, and design challenges. In addition, it provides a reader's guide.

• **Chapter 2 Related work and Background:** section one provides the relevant work to our framework. Section two represents a background of used techniques.

• **Chapter 3 proposed framework architecture:** In this chapter we illustrate how we achieved the framework design objectives.

• **Chapter 4 Evaluation and Experimental Analysis:** this chapter shows the experimental design, several experiments, case study and discussion.

• **Chapter 5 Conclusion and Future Work:** The chapter concludes the results that we have achieved and illustrates the applicability of the developed framework. Finally, it presents some work that can be investigated in the future.

# CHAPTER 2

# RELATED WORK AND BACKGROUND

The development of the proposed framework has involved review of research in various computer science fields. This chapter is oriented towards three main subjects: (1) Efficient management of the network scarce bandwidth resources using different techniques such as progressive mesh streaming. (2) Development of light weight applications. (3) Caching mechanisms that reduce the amount of data requested over the network and improve the response time.

## 2  Related Work and Background

### 2.1  Related Work

Reducing data transmission in the wireless medium is essential to improve any virtual environment system. This can be achieved by various approaches such as:

- Creating and developing efficient models for storing, transmitting, and displaying 3D objects in multimedia and VR applications [4] [5] [6] [7] [8] [9] [10] [11] [12].
- Designing special network transmission protocols that suit new generations of mobiles [13] [14].
- Pervasive mobile computing and cloud computing (light-weight applications) to overcome processing power limitation [15] [16] [17] [18] [19] [20].
- Devising data caching techniques on device cache memory. [1] [2] [17] [21] [11] [22].

Progressive mesh techniques (PM) are used to simplify 3D objects and beat the limitation of wireless network bandwidth [23] [9]. Authors in [11] presented a progressive mesh representation method to simplify the complex triangle mesh into a coarse one. The mobile users can get a general idea about the 3D model without long waiting time. An explicit rule has been used to calculate the cost of each collapsing edge.

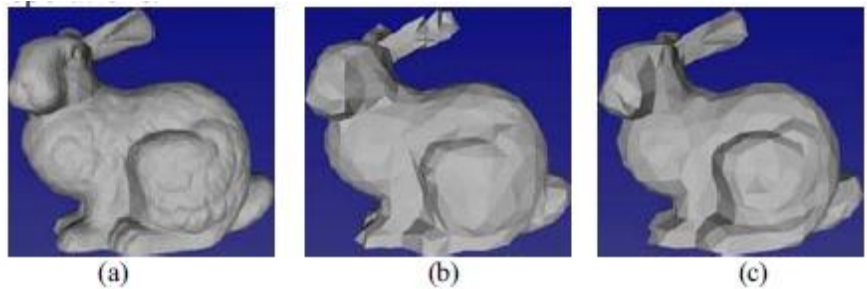

**Figure 2: representation results of different algorithms. (a) Original Bunny with 9999faces. (b) Base mesh using PM of quadric error measuring (c) base mesh using [11] method**

Figure 2 represents a test result of bunny model using progressive mesh representation method and quadric error measuring. The base mesh contained about 1000 faces. The base mesh was much more approximating the original one in their method than another. However, this approach lacks the view-dependent ways to represent the whole mesh.

J.Ma et al [6] presented a 3D mesh compression approach based on reverse Modified Loop scheme and embedded zero-tree mesh coding. The dense mesh was decomposed into progressive mesh which can be flexibly transmitted over the wireless network.

Table 1: Compression Ratio by file size [6]

| Model | Verts. | Geo. (KB) | Con. (KB) | Total (KB) | Orig. (KB) | Comp. Ratio (%) |
|---|---|---|---|---|---|---|
| random | 4,338 | 13.65 | 0.80 | 14.45 | 444 | 3.25 |
| egea_u | 5,315 | 14.90 | 0.60 | 15.5 | 540 | 2.87 |
| fandisk | 6,475 | 12.69 | 1.64 | 14.33 | 520 | 2.76 |
| venus | 8,268 | 23.05 | 3.53 | 26.58 | 800 | 3.32 |
| foot | 10,016 | 24.06 | 3.75 | 27.81 | 945 | 2.94 |
| sphere | 10,242 | 28.36 | 0.10 | 28.46 | 1,030 | 2.76 |
| tf2 | 14,169 | 31.90 | 2.91 | 34.81 | 1,170 | 2.98 |
| horse | 19,851 | 44.62 | 4.19 | 48.81 | 2,070 | 2.36 |
| maxplanck | 25,445 | 58.39 | 9.29 | 67.68 | 2,630 | 2.57 |
| feline | 49,864 | 103.00 | 14.99 | 117.99 | 5,000 | 2.36 |

Table 1 represents the compression ratio by file size. The original size is listed in the 6th column. The total of compressed files is in the 5th column with the ratios in the $7^{th}$ column.

Several derivatives of the progressive mesh techniques are existe in [7] [9] [5]. Isenburg and Lindstrom [7] proposed the streaming meshes technique where a mesh is stored into a fixed size buffer and triangles and vertices are either added or removed from the mesh in order to reconstruct it. Kircher and Garland [9] proposed a multi-resolution representation for deforming objects with a high quality approximation. The multilevel mesh proposed by the authors aims at having iterative edge contraction, and uses less space since it stores progressive representation (mesh connectivity at each level) instead of the entire hierarchy. However, edge contraction makes the mapping false since vertices forming the children can move [5]. Fang and Tian [10] implemented a mesh simplification based on the triangle contraction simplification. Pajarola and Rossignac [8] proposed the compressed progressive meshes (CPM) approach aiming at improving the PM technique by focusing on removing the overhead and latency engendered by the progressive mesh. For this matter, CPM uses the implant sprays technique to refine the mesh by assembling the vertex splits into batches. In consequence, CPM occupies 50 percent less storage than the PM model. Modified Compressed Progressive Meshes (MCPM) technique [4] improves CPM by including a decision module that selects the most suitable transport protocol for each geometric sub-layer taking into account the network bandwidth and the loss ratio.

Chi-Kang Kao et al [24], proposed a multi-resolution representation of 3D animation that resulted in displaying 3D animation progressively. The framework used to compress and to simplify the original animation into a smaller, lower polygon form, and allowed progressive visualization of the animation.

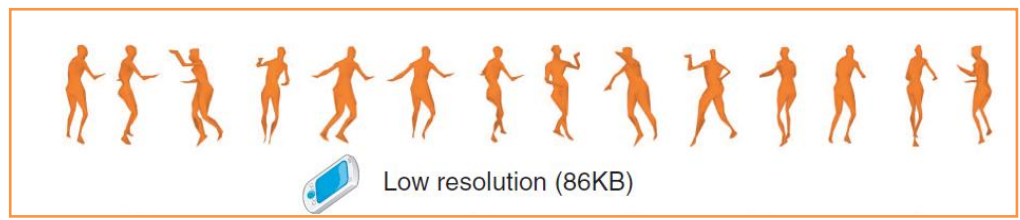

Figure 3: Progressive animation for low resolution [24]

The proposed framework for progressive animation was achieved by constructing the PCR (progressive connectivity representation) and PGR (progressive geometry representation) on the encoding side. KPCA (K means +Principal component analysis) is used to compress the geometry data into a more compact format called PGR. The PGR supported refining each vertex trajectory individually by utilizing the property of PCs. The PCR represented a progressive connectivity in terms of the base mesh with prioritized Vsplit. The PCR handled the sequences of split vertices and connectivity updating (CU), and the PGR maintained the coordinates of subset vertices and provides geometry updating (GU) to the vertices with different required sizes.

A. Boukerch et al [15] proposed a remote walkthrough over a heterogeneous network for PDA devices using Image Base Rendering (IBR).

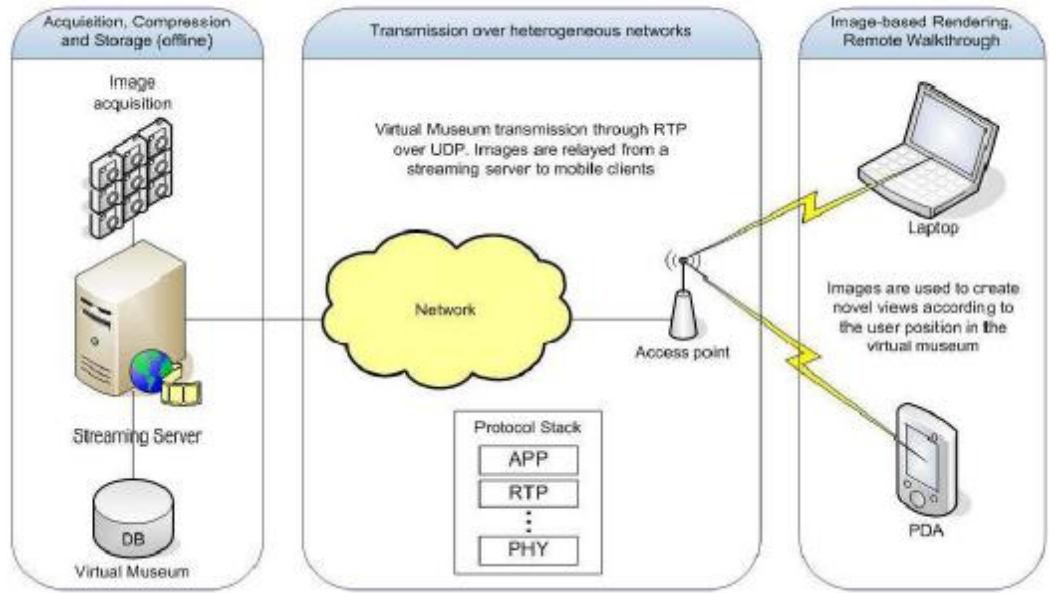

Figure 4: Remote walkthrough system [15]

In 2008 they used end-to-end virtual environment streaming technique [14].

Recently published researches start to develop light weight algorithms, protocols and applications for mobile devices to improve the applications efficiency and performance [19] [25].

A. Boukerche et al [**15**] presented a client/server approach, which consists of payload format for round trip time (RTP), an interactive streaming algorithm, the packetization scheme, and the pre-fetching mechanism for remote interaction in virtual 3D environments for mobile devices over ad hoc networks. View morphing technique has been used.

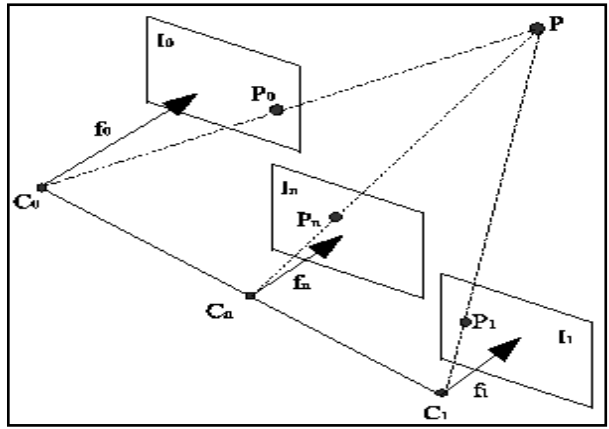

Figure 5: View morphing with parallel views [15]

View Morphing was used to render any image by morphing two or more reference images. However, the disadvantages of View Morphing technique are speed and control. Because it is global, all line segments need to be referenced for every pixel, which can cause speed problems. A location-based mobile tourism application using a cloud-based platform to reduce the power consumption in mobile device was used in [**26**]. The architecture presented a location-based mobile tourism application for iOS platforms as a front-end level. Web service is used to generate XML output from database Amazon MYSQL RDS to exchange data with mobile applications in middle-ware level. Servers were built in the cloud by using Amazon Web Services cloud platform. A similar study was conducted by Bao-Shuh P. Lin et al [**16**], in which, they used cloud-based 4G broadband mobile system (BMS) (TD-LTE) to develop mobile augmented reality (MAR) applications. They addressed a design of navigation/tourism applications for indoor and outdoor, collaborative urban design, and multiuser interactive motion learning system.

Caching techniques are widely used in different applications to reduce the amount of data transferred through the network and to provide a better quality of services [**27**].

In [**28**], the authors took the replacement decision based on the information of the avatar's geometric attributes. They used three levels Level of details (LOD) to present the models. All models on the screen are rendered according to their priority in prefetching and cache replacement. More detailed model indicates higher prior.

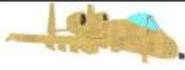

Figure 6: Illumination of different LoD [28]

They used integrated information, including geometric attributes and non-geometric attributes. This technique proposed prefetching and caching scheme according to the integrated information. Caching and prefetching were based on the Degree of Attention about Object's Attributes (DAOA) technique [**28**]. In this method the avatars can dynamically decide which object is concerned with mapping its information of attention to another object's attributes, and determine priorities of objects whose 3D geometries need to be prefetched or cached. Also, the user can change his degree of attention easily in the processing of distributed virtual reality (DVR).

A. Boukerchet et al [**15**], presented mechanism for Guntella network to support mobile collaborative virtual environments (CVE) application over ad-hoc network. They used caching technique to control the node flow and the amount of data exchange between nodes. The authors generate pong caching mechanism that could be implemented in any mobile Gnutella client in order to support mobile CVE applications. Each entry stores information depending on the application such as IP address, port number, time stamp and so on. The mechanism used a separate data table for each connection; the refresh process was also controlled separately. Whenever the data table size fell below a threshold value, the refresh process was started only via the connections having a small data table size. Initially all data tables contain no pong cache data. The choice of pong cache data that should be returned to the originator should meet predefined PING conditions, nodes should have a minimum hop value with a better CVE-profile matching, which was defined as the same session ID and a maximum number of zones that overlap with the zone interests of the originator node. Each node is requested to maintain and monitor its cached data table in order to remove any outdated entry.

In [**17**] caching is used to store images on a local mobile device in the form of SQLite database to perform offline tasks.

The authors in [**11**] addressed the caching and prefetching technique called MLM (Most Likelihood movement). This model is used for caching and prefetching possible future objects on the mobile based on the user's projected intention.

This variety of approaches to develop efficient mobile applications is an expression of the multitude of applications domain to which mobile devices can serve. Table 2 summarizes the main characteristics of the most relevant publications to our proposed framework.

**Table 2: The Relevant Related work**

| Authors Name | Caching Technique | Rendering Techniques | Platform | Network | Architecture |
|---|---|---|---|---|---|
| J. Ma, Q.Chen, B.Chen, H.Wang [6] | ----- | Progressive Mesh | Mobile | Wireless Network | Client- Server |
| A.Boukerche, R.W. N. Pazzi [13] | Using Mobile Cache | Image base Rendering (IBR) | Mobile | Heterogeneous Networks | Client - Server |
| A. Boukerche, R. W.N. Pazzi , J. Feng, [14] | Using Mobile Cache | Image base Rendering (IBR) | Mobile | End –to- end Wireless network | Client -Server |
| A. Boukerche, R. W. N. Pazzi, T.Huang [29] | Using Mobile cache | Image base Rendering (IBR) | Mobile | Wireless Network | Client -Server |
| J. Chim, R. W. H. Lau, W. Cyber Walk [2] | Using Cache MRM Technique | Progressive Mesh | Desktop | Wired LAN | Client -server |
| R. W. H. LAU, J. P. CHIM, M. GREEN H.VA LEONG, A. SI, [21] | Using Cache MRM Technique | Progressive Mesh | Desktop | Wired LAN | Client -Server |
| H. Hoppe [1] [30] | ----- | Progressive mesh | Desktop | ----- | ----- |

These approaches are categorized according to the type of platform, caching techniques, rendering algorithm and network architecture. For example, the authors in [6] proposed a compression technique for progressive mesh transmission of one object but they didn't use any caching technique. While the authors in [15] [13] [14], proposed a virtual environment framework under heterogeneous mobile networks, they used IBR technique to display the scene. This rendering approach is visually annoying, with effects such as globally blurred images, artefacts at edges, loss of precision in texture rendering and detail coding. In our framework, we used progressive Mesh technique to avoid IBR visual deficiencies. Moreover, we execute all mesh processing tasks in the server side to keep the application deployed to the mobile as light as possible.

## 2.2 Background

In this section we will summarize the approaches used in the proposed framework

### 2.2.1 Progressive mesh

Geometric levels of detail (LOD) have been used in two forms for fast display of large environments: static LODs [31] and view-dependent simplification such as the progressive mesh [1] [30]. Progressive mesh (PM) streaming enables users to view 3D meshes over the network with increasing level of details, by initially sending coarse version of the object called base mesh, followed by a series of refinements, called Vsplits [32].

In the PM representation [1], an arbitrary mesh $\hat{M}$ is simplified through a sequence of *n edge collapse* (*ecol*) transformations as shown in figure 7 to yield a much simpler base mesh $M^0$

$$(\hat{M} = M^n) \xrightarrow{ecoln-1} \dots\dots\dots \xrightarrow{ecol\ 1} M^1 \xrightarrow{ecol\ 0} M^0$$

Because each *ecol* has an inverse, called a *vertex split* transformation, the process can be reversed:

$$M^0 \xrightarrow{Vsplit\ 0} M^1 \xrightarrow{Vsplit\ 1} \dots\dots\dots \xrightarrow{Vsplitn-1} (M^n = \hat{M})$$

The tuple ($M^0$, { $Vsplit_0$,……,$Vsplit_{n-1}$ }) forms a PM representation of $\hat{M}$. Each vertex split, parameterized as $Visplits(v_s, v_l, v_r, v_t, f_l, f_r)$, modifies the mesh by introducing one new vertex $v_t$ and two new faces $f_l = \{v_s, v_t, v_l\}$ and $f_r = \{v_s, v_r, v_t\}$ as shown in Figure 7.

Edge collapse is used for reducing the model resolution, whereas splitting is used for increasing the model resolution as illustrated in Figure 7.

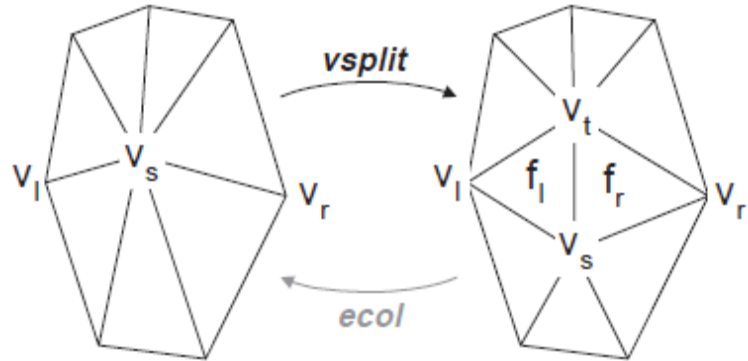

Figure 7: definitions of refinement (Vsplits) and coarsening (ecol) transformation [30]

Each object is modelled as an ordered list of resolution records starting with a minimal resolution model of the object (base mesh). Each subsequent record in the list is referred to as a progressive record; it stores information of an edge split [33]. The structure of a progressive mesh is shown in Figure 8.

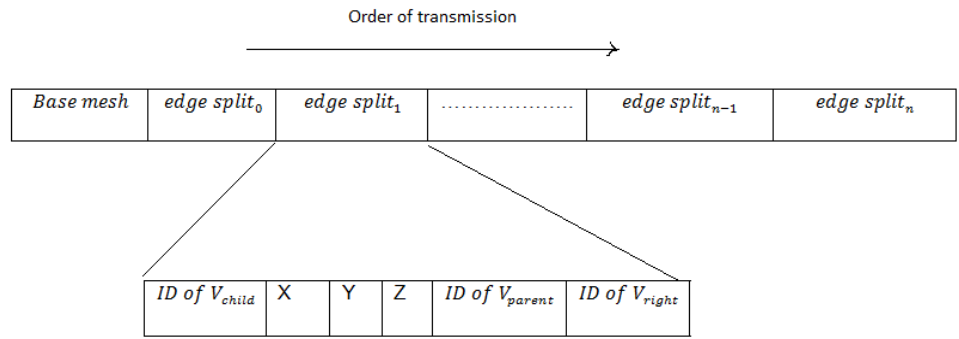

Figure 8: Data structure of progressive mesh for transmission [10]

If each progressive record is orderly applied to the base mesh, the object resolution will be gradually increased until it reaches the highest level of details.

Our proposed framework exploits mobile phones display limitations and further optimizes bandwidth utilization and power consumption by restricting clients' resolution requests to the maximum HW display specifications. Client side constrained object's resolution selection is used to identify the highest resolution of object's mesh that will be transmitted, based on the device capabilities such as the screen size and network bandwidth. And hence, avoiding the transmission and processing of unnecessary data.

### 2.2.2 Caching techniques

#### 2.2.2.1 Caching overview

The cache is a smaller, faster memory which stores copies of the data from frequently used main memory locations. When the processor needs to read from or write to a location in main memory, it first checks whether a copy of that data is in the cache. If so, the processor immediately reads from or writes to the cache, which is much faster than reading from or writing to main memory. The proportion of accesses that result in a cache hit is known as the hit rate, and can be a measure of the effectiveness of the cache for a given program or algorithm.

The heuristic that it uses to choose the entry to evict is called the replacement. There are many replacement caching techniques used to allow the new data to enter the cache memory; for example the CPU caching techniques [**34**].

***The Not recently used (NRU),*** sometimes known as the Least Recently Used (LRU), page replacement algorithm is an algorithm that favours [**35**] keeping pages in memory that have been recently used.

***First-in, first-out,*** the simplest page-replacement algorithm is a FIFO algorithm with low-overhead as it requires little book-keeping on the part of the operating system.

***RAND (Random)*** chooses any page to replace at random.

*LRU (Least Recently Used)* chooses the page that was last referenced the longest time ago.

*The Not frequently used (NFU)* page replacement algorithm requires a counter, and every page has one counter of its own which is initially set to 0. At each clock interval, all pages that have been referenced within that interval will have their counter incremented by 1. In effect, the counter keeps track of how frequently a page has been used. Thus, the page with the lowest counter can be swapped out when necessary.

### 2.2.2.2 Most Required movement Caching Technique

There are several caching techniques proposed in the literature. The proposed framework employs the Most Required Movement (MRM) replacement technique to define an access score for each object [**2**]. The scheme is based on the observation that the farther an object is from the viewer, the lower the resolution it can be rendered at, and the longer it will take the viewer to move to a point where greater details are required. Similarly, the larger the angle between an object and the viewer's line of sight, the lower the resolution required and the longer it will take a client to rotate to directly view the object [**21**]. The probabilities of being rendered at a higher resolution and of being cached in the storage are also lower. When the object with the lowest access score is selected for replacement, the object is not totally removed from the cache. Rather, its extra resolution details are reduced till it reaches its minimal resolution (the base mesh), this makes room for the new incoming objects. If there is no enough space for accommodating new objects, then the object with the next lowest access score will be replaced.

# CHAPTER 3

# PROPOSED FRAMEWORK

This chapter describes the framework methodology and its components in details. The chapter is organized as the following: section 3.1 the framework scenario overview, section 3.2 describes the framework components and section 3.3 shows the framework sequence diagram.

# 3 The Proposed Solution

## 3.1 Applying framework on mobile device

Mobile device still lack the proper resources to run graphic intensive applications such as walkthrough applications. The proposed framework tries to solve this limitation and generate light weight application suitable for mobile device.

As we mention in previous chapter several techniques have been used to apply walkthrough application on mobile device. Some challenges are faced to develope our framework on mobile device such as:

Progressive mesh is used to virtualize 3D objects of low quality and then refine them if needed the first challenges in progressive mesh on mobile is it takes a long time for generating the mesh. The problem solved by creating the mesh partition in server side this create light application which clients just render the 3D object on mobile screen.

The second challenge in progressive mesh is it generates a large number of records this leads to congestion on traffic. , constrained object's resolution solve this problem by selecting module is used to identify the highest resolution of the object's mesh that will be transmitted, based on the device capabilities such as the screen size and network bandwidth. Thus, it avoids the transmission and processing of unnecessary data.

According to the limitation on bandwidth in wireless network using mobile device, we reduce the progressive mesh records by dividing 3D object to 10 records.

Caching techniques are widely used to reduce the amount of data transferred through the network. In our proposed framework for walkthrough applications, we store, update and replace the objects in the device cache according to the distance between the user's viewpoint and the object, in order to provide a better quality of service to the users. This approach showed a great efficiency in walkthrough applications in [2] [21]. We implemented the Most Required Movement (MRM) replacement technique on mobile device and showed its significant impact on the performance in our experimental result. Next section explains in details our framework and its components.

## 3.2 The Framework Scenario Overview

The developed framework is Client-Server architecture. The simulation life cycle is shown in figure 9.

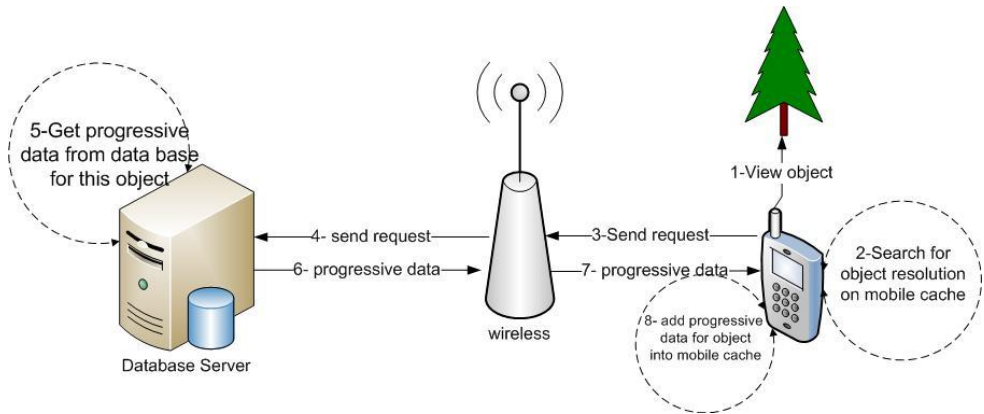

Figure 9: Framework life cycle

The user starts to navigate through a virtual scene. Before requesting the object's resolution from a database server, the client searches for the progressive records in his cache. If it is not found, it sends requests to the database server through the wireless medium. The server sends progressive records back according to the mobile's configuration such as the screen size and bandwidth that were stored on its database. Then, the user receives progressive records from the server and puts them into his cache.

## 3.3 Framework Main Components

The proposed framework is divided into four main modules: the client module, the wireless medium (Uplink, downlink) modules and the server management modules as shown in Figure 10.

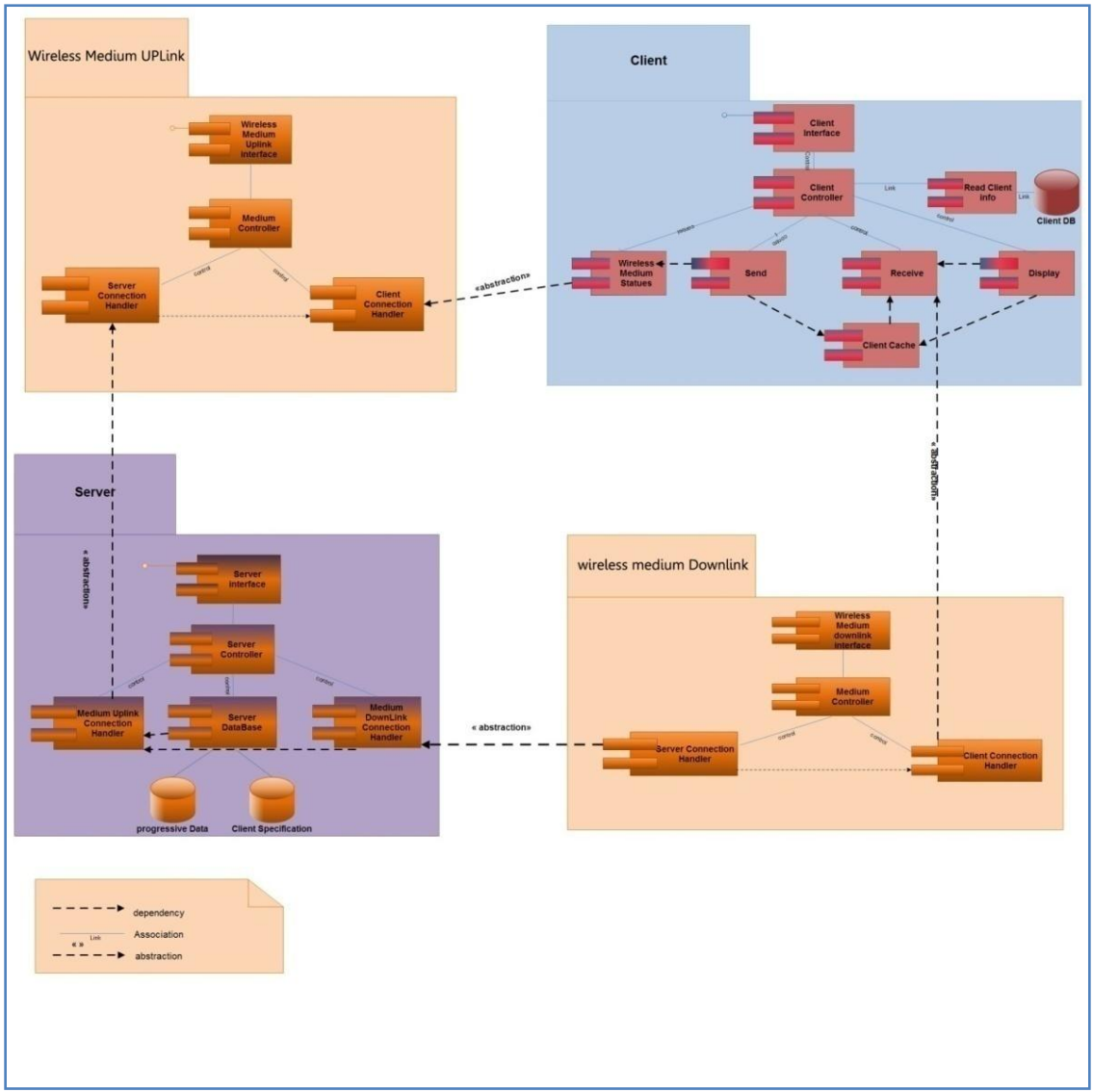

**Figure 10: Framework components**

### 3.3.1 Client subsystem

The client module simulates the mobile device side and controls navigation, rendering, communication, and cache memory. The user starts to navigate through the scene and the scene objects are progressively received in multi resolution mesh format as shown in figure 11.

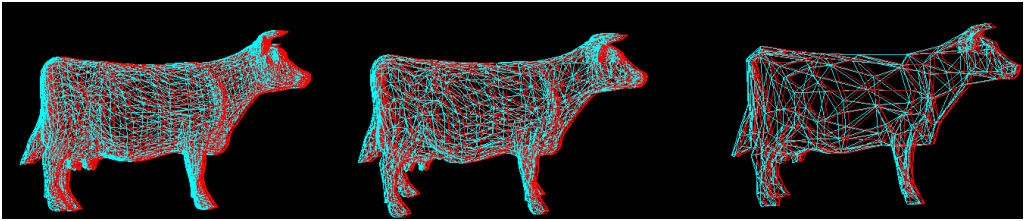

Figure 11: An example of 3D object in progressive format (from left to the right) 100%, 50% and 10%

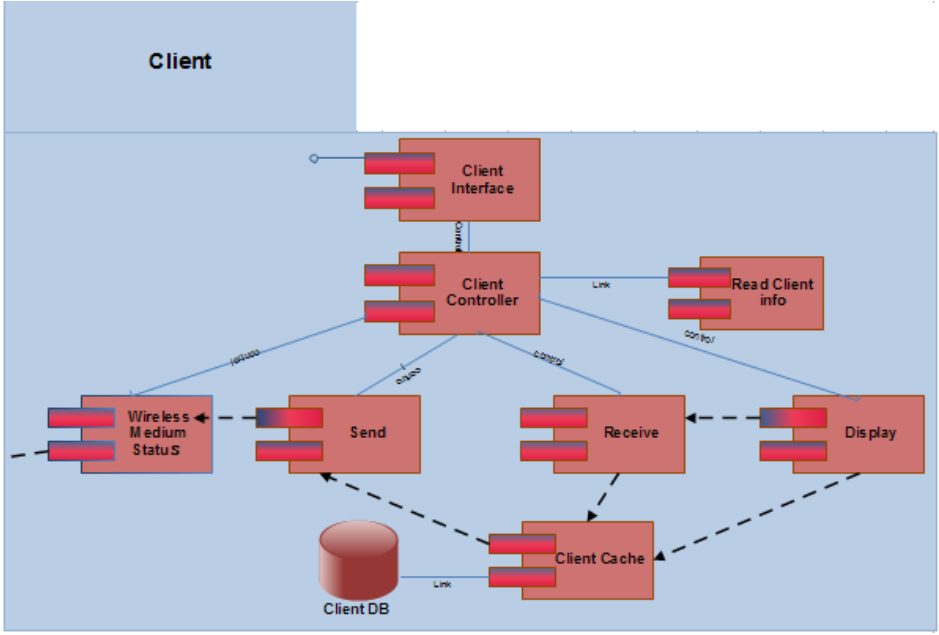

Figure 12: The client's module components diagram

The client module started with client interface which has association relation with the client's controller as shown in figure 12.

***Client controller*:** it coordinates all other element components in the client's module. It also simulates user navigations and prepares objects resolution requests to be sent to the server by the communication agent. The client controller element reads the client's info from the client's database and starts up all element components into this module.

***Wireless Medium Status*:** this element is responsible for checking the medium status (busy or free) to know the availability of send request to the server through the wireless medium.

***Send:*** in charge of sending requests from the client to the wireless medium up link. This element has dependency relation with the wireless medium status so it can't send any request until the wireless medium status element report that the wireless medium is free.

***Receive:*** it is responsible for receiving the progressive data that is sent from the server through the wireless medium downlink and storing the coming data into the client's cache.

***Display*:** this element is in charge of rendering progressive data on the mobile device. It has dependency relation with received element so it can't display until the progressive data is received from the server and also has dependency relation with the client's cache.

***Client cache:*** is responsible for all operations that happen on the client's cache such as search for object, add object and replace object from the cache. The client cache has dependency relation with the send element because the element send can't send requests to the server until the client's cache search for object in the cache and report if the requested object found or not. Client cache also has dependency relation with receive element because the client cache can't put the incoming data into cache until receive element report that it is received from server.

## 3.3.2 Wireless Medium (Uplink & Downlink)

The wireless medium modules are responsible for sending requests and data between the client module and the server module. CSMA/CA protocol is simulated to handle data transmission [3].

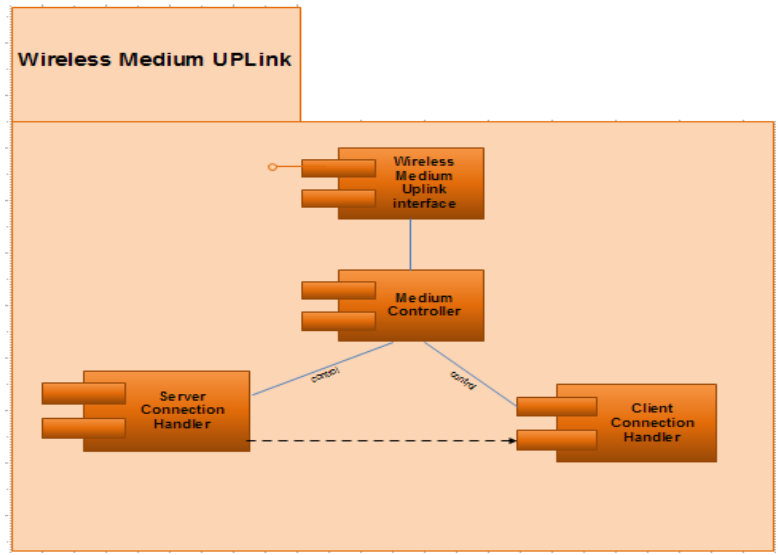

Figure 13: wireless medium uplink

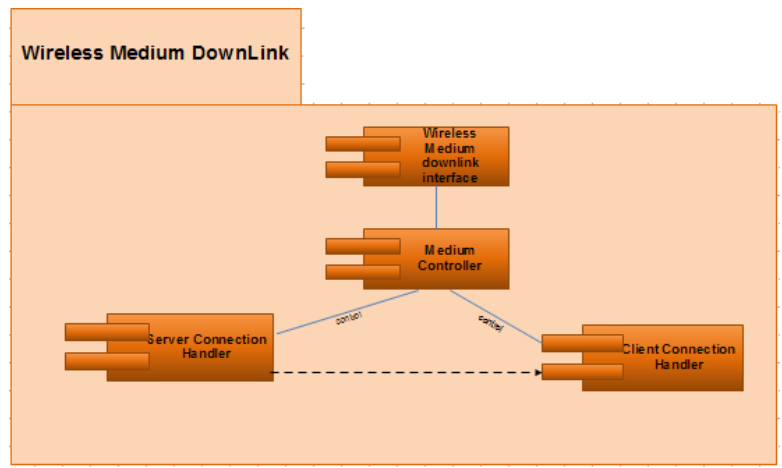

Figure 14: wireless medium downlink module

Both of the wireless medium up link and downlink has an interface that has association relation with the medium controller as shown in figures 13 and 14.

*Medium controller***:** in both of the wireless medium uplink and downlink it is responsible to start all elements into the module.

*Client Connection Handler (Uplink):* this element is responsible for handling the connection between the client and the wireless medium up link. It sends broadcast message to all the clients that are connected to the framework containing his status (free or busy).

*Server Connection Handler (Uplink):* this element controlling the connection between the wireless medium up link and the Server. It is responsible for sending the client request to the server.

*Server Connection Handler (Downlink):* it opens connection between the server and the wireless medium downlink.

*Client Connection Handler (Downlink):* it controls the connection between the wireless medium downlink and the client to send the progressive data from the server to the clients.

### 3.3.3 Server Module

The server manager module runs on the server side. It consists of five agents and two databases:

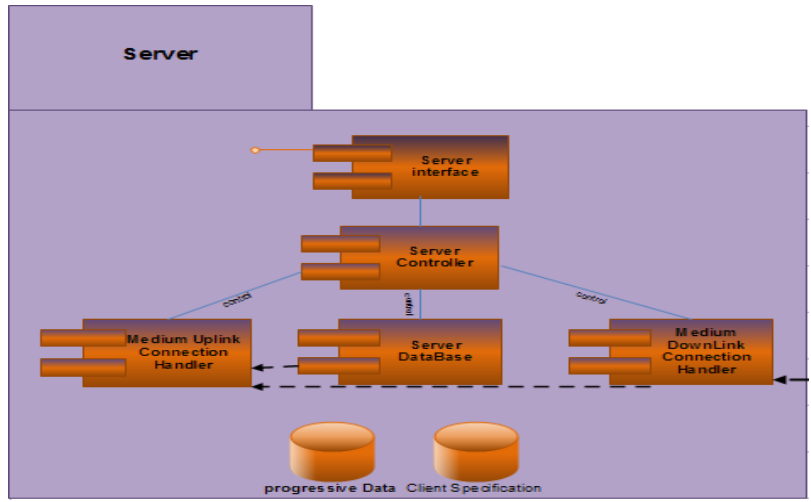

**Figure 15: Server Model**

The Server module consists of the Server interface that is connected with the server controller element with associated relation as shown in figure 15.

*Server Control*: is responsible for controlling and starting all elements in the module.

*Medium uplink Connection Handler*: it opens and controls the connection between the server and the wireless medium up link to get the client requests from the client module.

*Server Database*: is connected with the server databases that contain all 3d objects in progressive mesh representation and the database that contains the client specification. The server database has dependency relation with the medium up link connection handler. It can't get the object data from the data base until the wireless medium up link sends the client request to the server.

*Medium downlink connection Handler:* it controls the connection between the server and the wireless medium downlink to send progressive data from the server to the client through wireless medium down link.

## 3.4   Sequence Diagram

Figure 16 represents the sequence diagram of the proposed framework that contains five instances: client, client's cache, wireless medium up link, server and wireless medium downlink.

The user navigates using a mobile device, the client instances read the mobile configurations and generate requests id. The wireless medium sends the current status to the client instance; if it's busy the client waits a period of time until the status change to free to be able to send his request. The medium up link transfers the client's request to the server. The server makes some checking then gets progressive data from the database and sends it to the medium down link that is responsible for transferring data to the client. When the request is received by the client, the client makes caching checks (add or replace) then the 3d objects are rendered on the user screen.

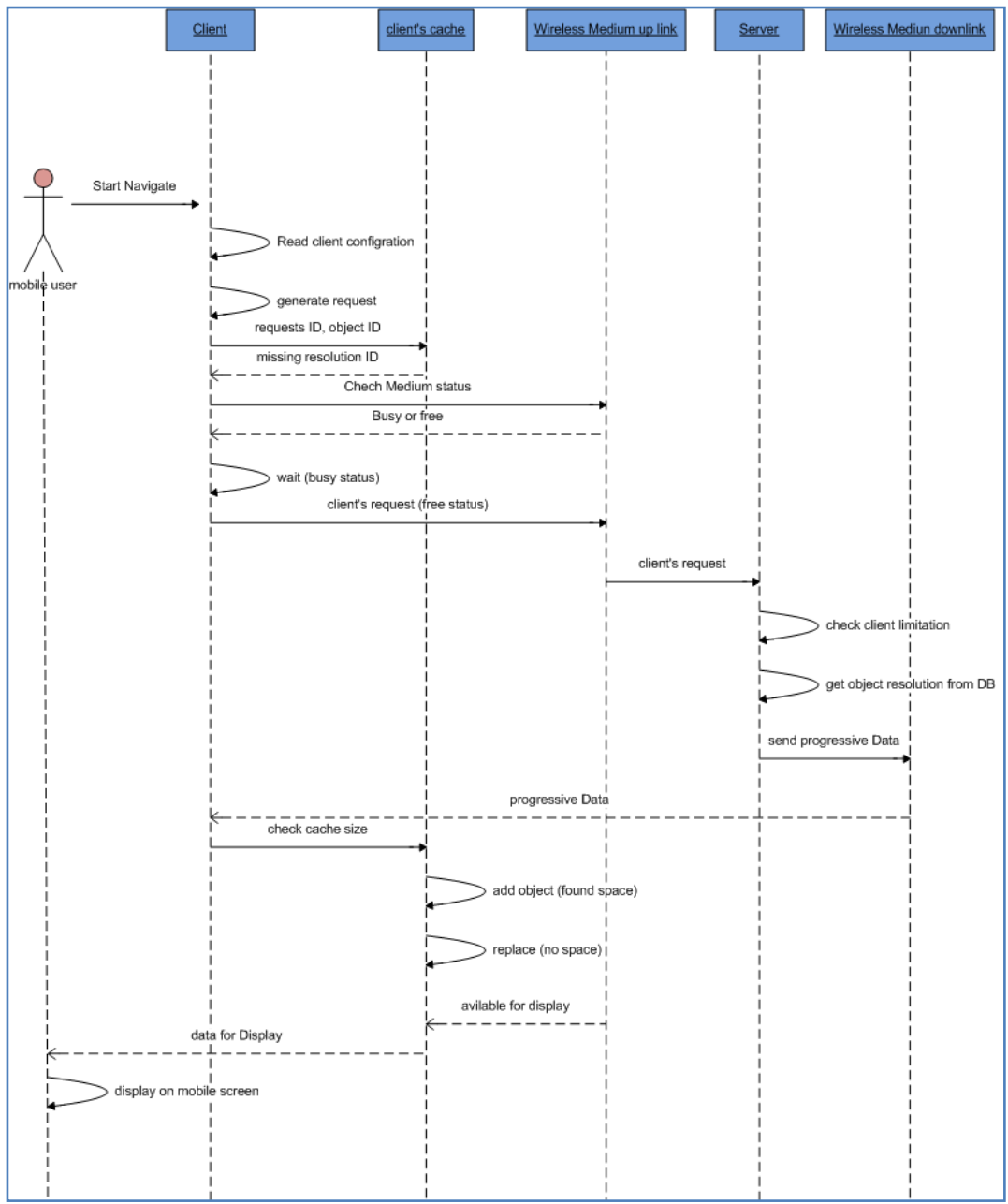

**Figure 16: Framework Sequence Diagram**

# CHAPTER 4

# EVALUATION AND EXPERIMENTAL ANALYSIS

This chapter conducts extensive experiments to quantify the performance of a virtual walkthrough on an interactive mobile application and on a variety of simulated and real environment. This is necessary for three reasons. First, a simulation environment will allow us to test the framework on different types of mobile. Second, to study a wide variety of parameters that affects the walkthrough performance. Third, a real world experiment is needed to prove that the framework is working efficiently on real mobiles and has the same performance trend as in the simulation environment. The chapter is organized as follows: section 4.1 explains the experimental Design, section 4.2 illustrates the experimental results, and finally section 4.3 includes our discussion.

# 4   Evaluation and Experimental Analysis

## 4.1   Experimental Design

The efficiency of the proposed framework is studied by measuring various virtual environment metrics. In the simulation, a virtual scene is created with 100 objects. Clients who navigate the scene follow: circular, change circular or random movement patterns. Each client has an area of interest (AOI) called viewer scope as shown in figure 17.

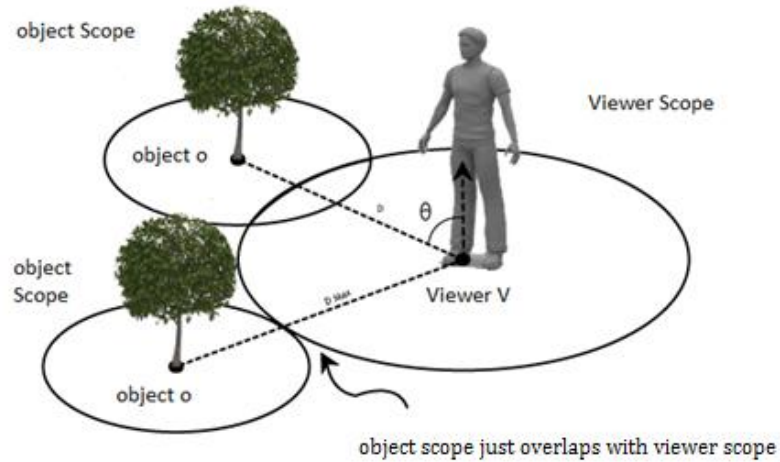

Figure 17: Interaction between the viewer and the surrounding objects

An object may be visible to a particular viewer only when the object scope overlaps with the viewer's scope [2]. Objects are transmitted and visualized in lower quality (base mesh) and then are progressively refined according to the distance from the client until they reach the optimal resolution.  It is the responsibility of the client module to render dynamic resolution objects based on the distance between the viewer and scene objects. The client sends requests to the server demanding object's resolution records only if they are not available in the local cache.  For example, if a client will render a 50% object resolution, then his request will be divided into 5 resolution requests (10%, 20%, 30%, 40%, and 50%) to simplify searching the cache memory for requested objects' resolutions. A memory efficient hash-map data structure is used to implement the local caching mechanism in the mobile devices. The cache structure is shown in figure 18.

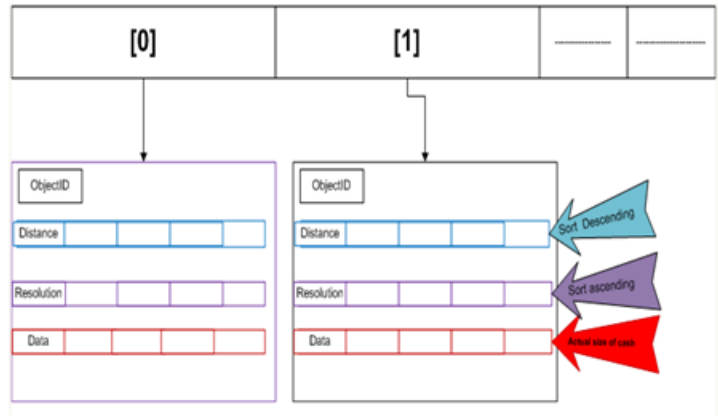

Figure 18 : Cache structure

The cache replacement policy is based on the assumption that a farthest object from the client will not be requested again in the near future. Hence, when free cache space is needed, the outermost object is selected as a victim. Then, the victim object's records are deleted starting from the highest resolution ones until reaching the base mesh. The procedure is repeated on the remaining objects if the freed space was not enough. This policy is similar to the widely used replacement policy called least recently used (LRU).

The scene objects are stored in a database as a sequence of progressive mesh records. The database application uses a dynamic mechanism to determine the number of records needed to render and visualize the objects in a certain position. The database is divided into three types of table; the first one contains the base mesh coordinates (x, y, z, object-ID), the second one contains the Vsplits Coordinates(x, y, z, object-ID, Resolution) and the third one contains aces indices (index1, index2, index3, object-ID, resolution). MYSQL is used as a database management system and Java SDK multi-threading programming language is used to build the simulation environment. The simulation uses two types of mobile; Type I and Type II. Type I includes mobile devices with high quality display that can render up to % 80 of object's resolutions. Type II includes mobile devices with low quality display that can render only the base mesh of object's resolution (10% of object resolution). The size of used cache is 2M. The object size is 60 KB. The base mesh record size is 8KB and the size of each resolution record is 5.2 KB.

## 4.2 Experimental Results

The simulation experiments evaluate the performance of the proposed framework using progressive mesh technique, constrained object's resolution request and cache technique. We used the typical performance measures as shown in table 3 [**14**] [**5**] [**2**].

Table 3: Experiment Metrics

| Metrics | Definitions |
| --- | --- |
| **Response Time** | Time amount between client's send request for object to be rendered to the time when optimal resolution of the object is available. |
| **Latency time** | Time amount between client's sent request for object to be rendered to the time when the base mesh of the object is available. |
| **Cache hit ratio** | The percentage of bytes of the object that could be retrieved from local cache memory. |
| **Virtual perception** | The relative degree (in percentage) of 3D objects quality experienced by the viewer just after the move. |
| **Network Utilization** | The ratio of wireless medium busy time to the total simulation time. |
| **Network Traffic** | The number of requests that are sent by the client to the server. |
| **Frame per Second (FPS)** | The number of frames rendered per second as a measure for display device performance |

In the following section, we explain the experiments that are carried out and discuss the results.

### 4.2.1 Experiment 1

The goal of this experiment is to evaluate the effect of using progressive mesh technique on the performance of the framework in terms of response time, latency time, virtual perception network traffic, wireless utilization and cache hit ratio. The experiment is conducted in the simulation environment using progressive mesh and is compared with the same obtained using static level of details. The setting of the experiment is shown in table 4.

Table 4: Experiment1 Parameters

| Parameter | Values |
|---|---|
| Number of clients | 20 clients |
| Cache size | 2 MB |
| Type of mobile | Type 1 |
| Static resolution | 50% and 100% resolution |

Figures 19 and 20 show the simulation average response time and latency time using both progressive mesh technique and static level of details' resolutions respectively.

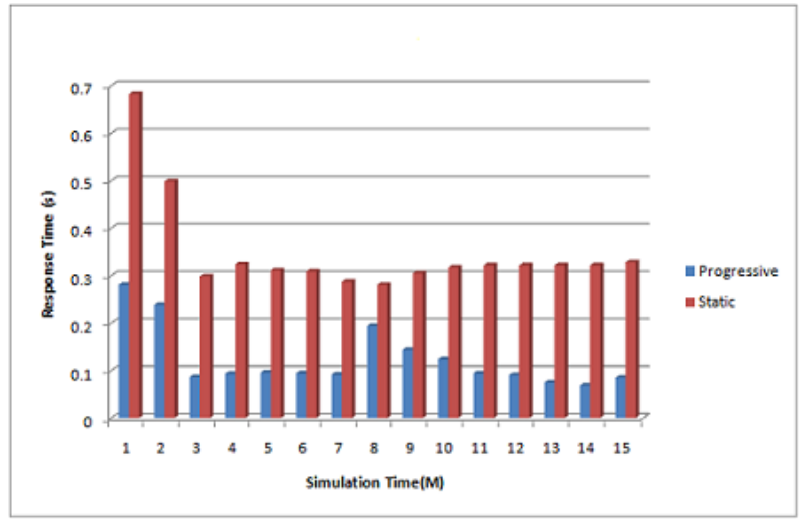

Figure 19 : Response time in progressive and static LOD

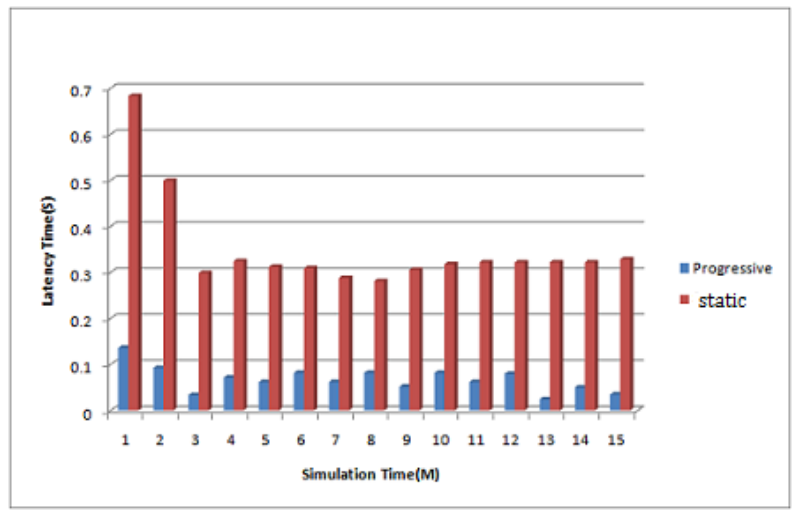

Figure 20: Latency time in progressive and Static LOD

We ran the simulation for 15 minutes and calculated the performance measures. As expected the response time and latency time in the static level of details took a long time. This is due to the transmission of large objects via a scarce wireless medium. In the first two minutes the simulation maintains high values for both of response time and latency time; this is due to the initialization time at the start up. After that, the response time varies according to the traffic of the object resolutions requested and how much of objects resolution are cached, as shown in figure 21.

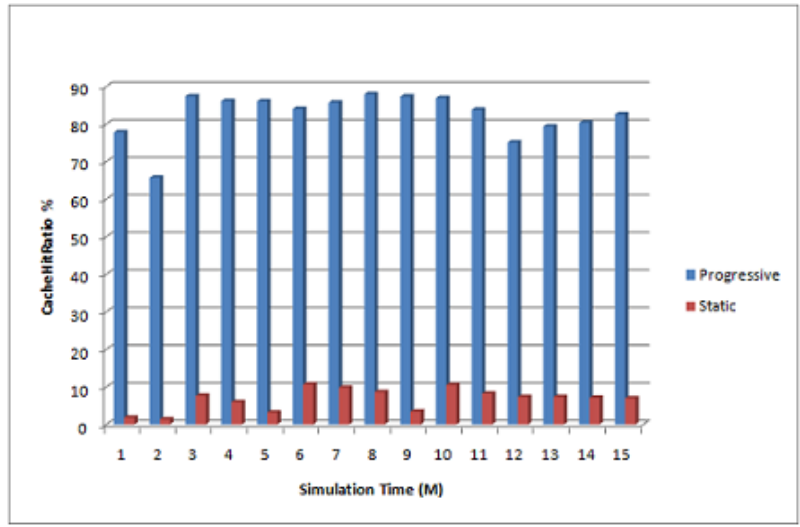

Figure 21:  cache hit ratio in progressive and Static LOD

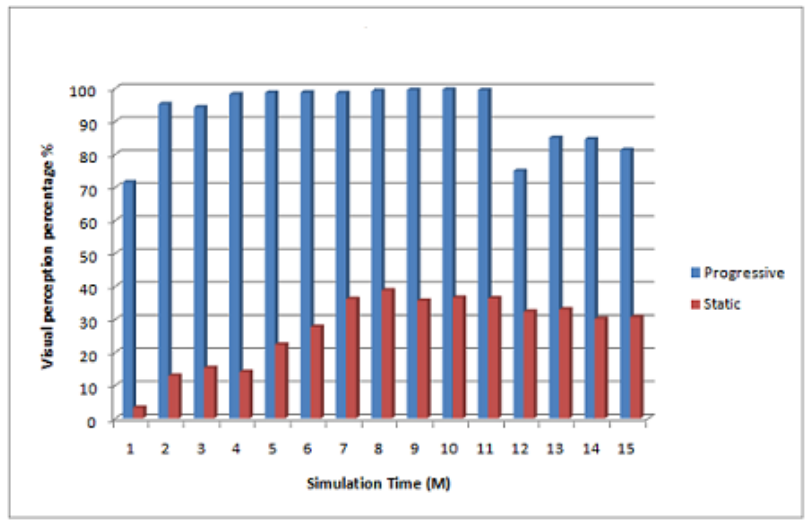

Figure 22:  Virtual perception in progressive and Static LOD

The virtual perception is one of the most important metrics for more interactivity in walkthrough applications. Figure 22 depicts the visual perception in both static level of details and progressive mesh resolution. The visual perception is represented by a cubic function: $1-(B_0 - B_0^*/B_0)^3$ where $B_0$ is the expected size of the object O at its optimal resolution and $B_0^*$ is the size of the object currently cached [2]. The perception metrics showed lower values with static LOD resolution, as the user had to wait for a long time before he gets his target object resolution. On the other hand, caching incremented records of resolution while waiting for a short time till the records are transmitted, resulted in better perception metrics and hence better user's virtual experience. The finer the granularity of resolution the faster the transmission of resolution records is, and hence the higher probability of caching large percentage of the objects resolution. This explains high values of visual perception when progressive mesh representation is used.

The progressive mesh divided the 3d object into small records size which leads to some advantages and disadvantages. The progressive mesh increased the number of requests and wireless utilization on uplink channel as depicted in figures 23 and 24.

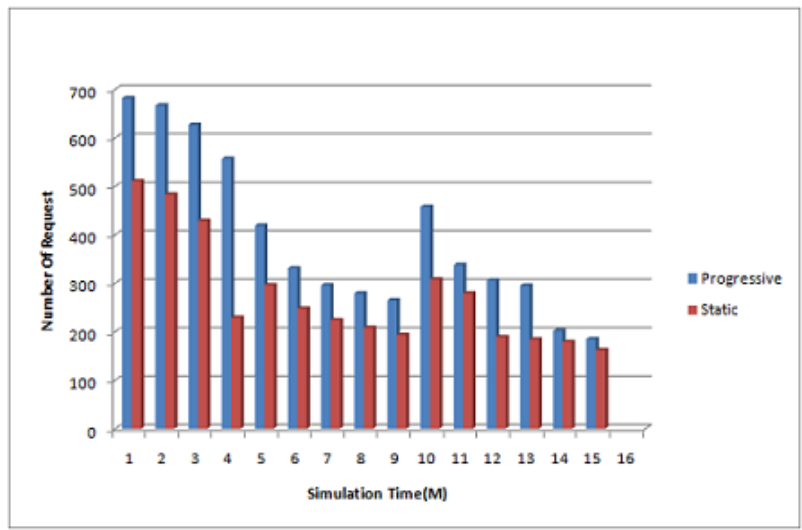

Figure 23 : Number of Requests in progressive and Static LOD

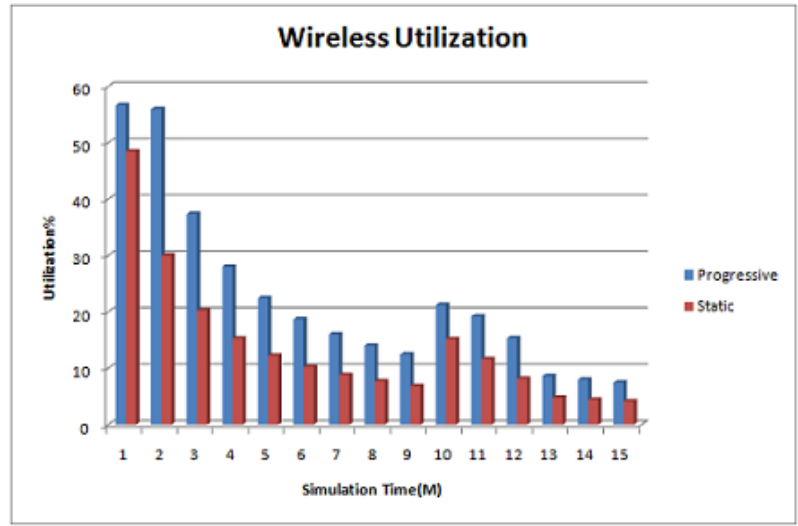

Figure 24: Wireless utilization in progressive and Static LOD

On the other hand, the down link medium utilization is improved when the server progressively transmits objects data to the client, as the medium is not needed to be utilized for a long time while it sends a large object at once. Instead, it will be released after short time periods of progressively sending small records one at a time, giving the opportunity to other clients' requests to be served and thus improving the overall users' perceptions. For objects whose sizes exceed the maximum package size of TCP/IP (11 MB [**12**]), segmentation couldn't be avoided and transmitting small chunks of object's data will solve the problem. Therefore, sending progressive records of an object is recommended for large objects in non-continuous network connectivity. Nevertheless, resending only a small part of a large object in case of packet loss is more bandwidth saving than resending the whole (large) object again.

### 4.2.2 Experiment 2

This experiment evaluates the effect of constrained object's resolution requests on the system performance in terms of response time, average number of requests and wireless medium utilization. The setting of the experiment is shown in table 5.

**Table 5: Experiment 2 Parameters**

| Parameter | Values |
|---|---|
| **Number of clients** | 20 client |
| **Cache size** | 0 MB |
| **Type of mobile** | Type II |

The simulation is run twice for two different sets of users. The first run simulates users with constrained object's resolution requests; while the second run simulates users with unconstrained object's resolution requests.

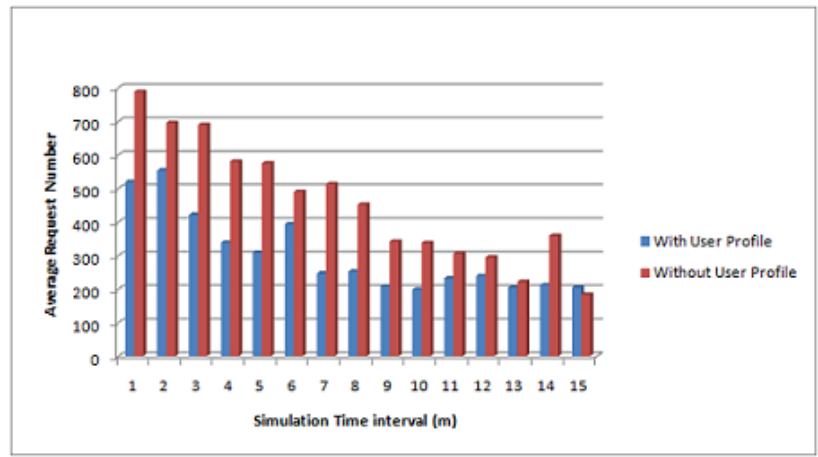

Figure 25: The impact of using constrained object's resolution request on traffic

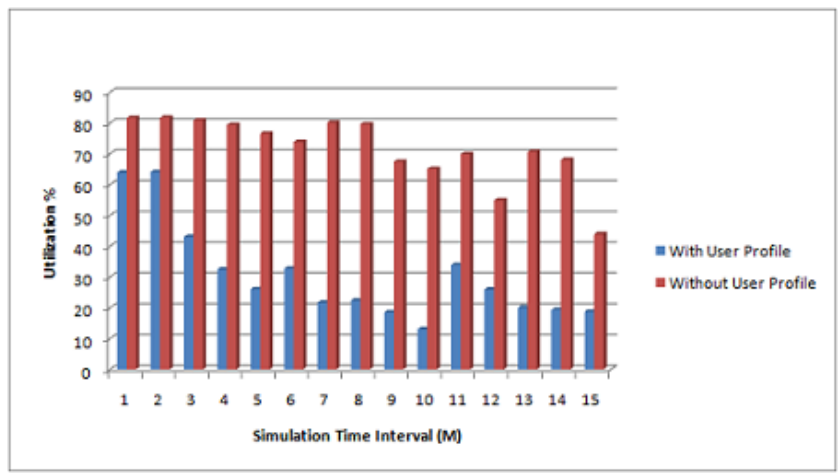

Figure 26: The impact of using constrained object's resolution request on wireless utilization

Figure 25 shows a decreased number of constrained objects resolution requests versus increased number of unconstrained object resolution requests. Figure 26 depicts small to medium wireless utilization values, between 10% and 30%, compared with large utilization values of 55% to 80% for constrained object's resolution requests and unconstrained object's resolution requests respectively.

Since the wireless medium utilization is proportional to the network traffic, it is expected that the response time will improve in low utilization network as shown in figure 27.

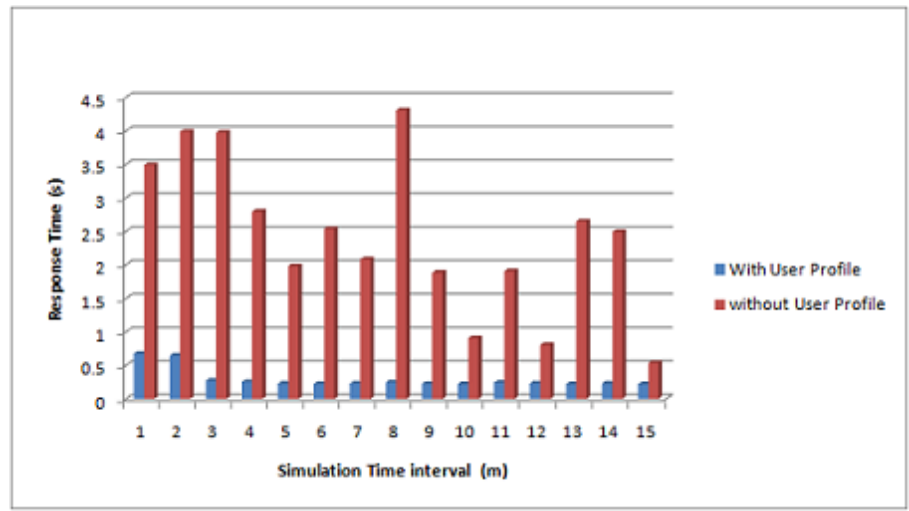

Figure 27: The impact of using constrained object's resolution request on response time

### 4.2.3 Experiment 3

This experiment evaluates the effect of using caching technique on the system performance measured in terms of average number of requests per minute, wireless medium utilization, and response time. Table 6 contains the experiment's settings.

Table 6: Experiment 3 Parameters

| Parameter | Values |
|---|---|
| **Number of clients** | 20 client |
| **Cache size** | Using 2MB and 0MB |
| **Type of mobile** | Type II |

The simulation runs twice with two different sets of users. The first set, uses mobile cache to store the current objects resolution records, while the second set doesn't use the mobile cache.

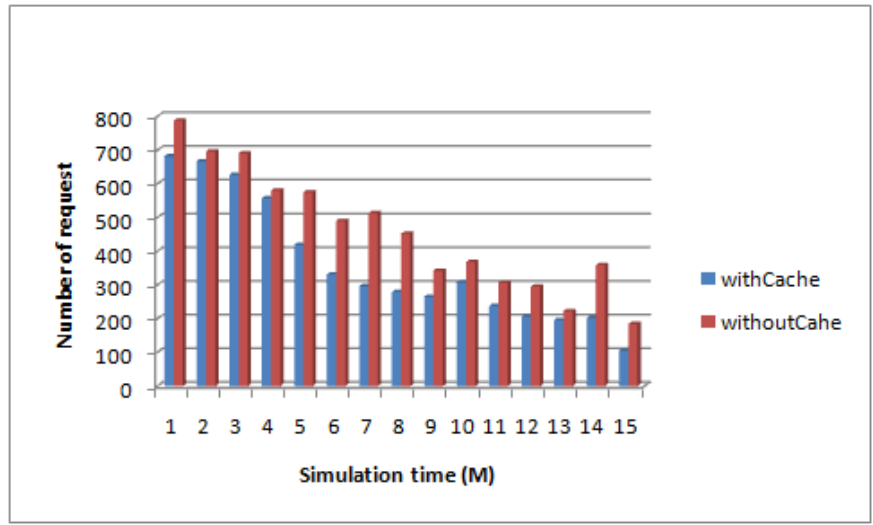

Figure 28: The effect of using cache on traffic

Figure 28 shows that the average number of requests per minute is decreased after using mobile cache. But after a while, the cache will start to get full and some saved objects resolution records need to be replaced upon addition of new ones. This pattern explains the sudden increase of number of requests after stabilizing for a while till new replacements are needed again.

That pattern is consistent with the graph in figure 29, as the wireless medium becomes busier with the number of requests increasing, and become less busy as it decreases.

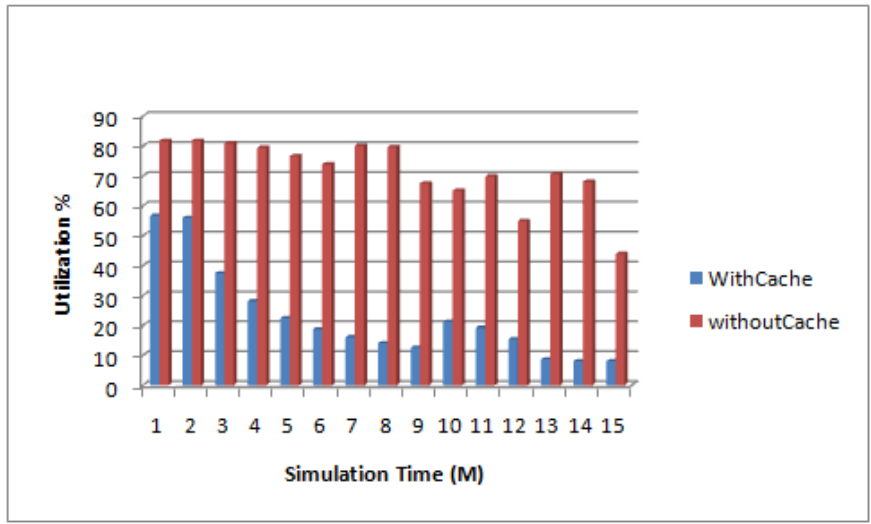

Figure 29: The effect of using cache on Wireless utilization

The response time is dramatically improved when using mobile cache as it is affected by both the network utilization and the requests sent to the server for retrieving resolution records, as shown in figure 30.

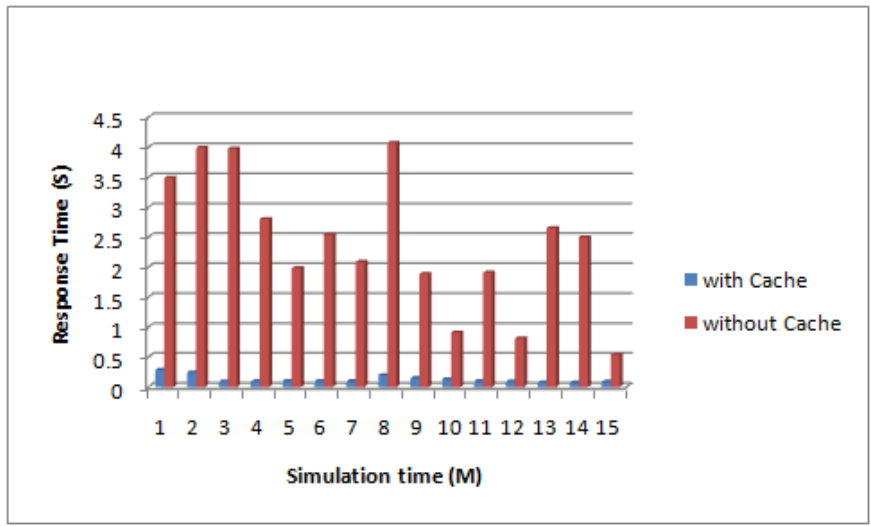

Figure 30: the effect of using cache in Response time

### 4.2.4 Experiment 4

This experiment represents the effect of different movement patterns on the cache-hit ratio. Table 7 contains the experiment settings.

Table 7: Experiment 4 Parameters

| Parameters | Values |
| --- | --- |
| **Number of clients** | 10, 20, 30 and 40 clients |
| **Type of mobile** | Type I & type II |
| **Movement patterns** | Circular pattern (CP), changing circular pattern (CCP) and Random walk pattern (RW) |

The movement patterns are shown in Figure 31.

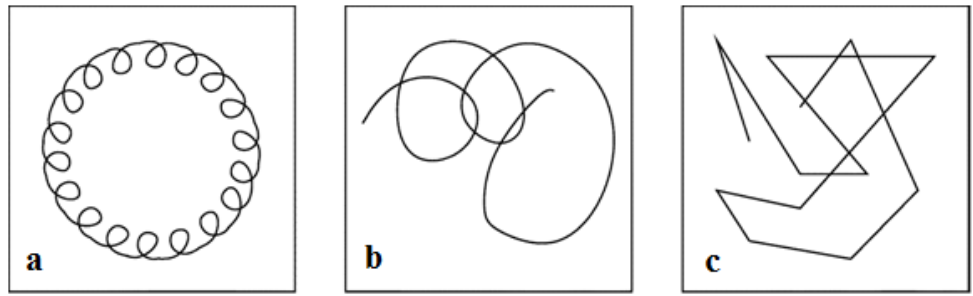

Figure 31: movement patterns (a) CP, (b) CCP, (c) RW [30]

Figures 32 and 33, display the cache hit ratio using the three movement patterns. As shown in figure 32, the cache hit ratio is higher with CP and CCP movement patterns than that with RW movement pattern for type I mobiles

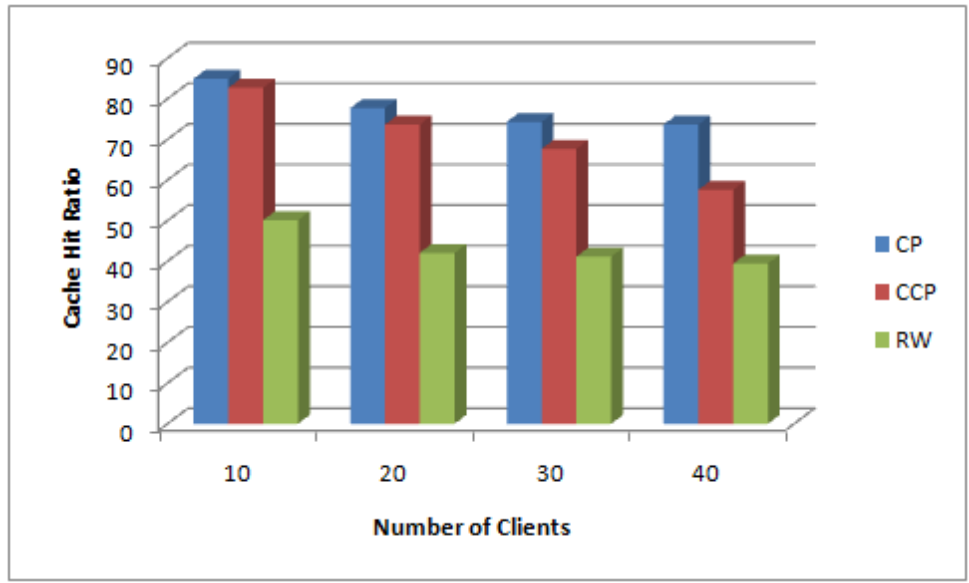

Figure 32 :    Average cache hit ratio for CP, CCP and RW in type I mobiles

This is mainly because the moving direction is always changing very often with a constant angle. Therefore, the probability to visit the same object is high. Thus, the device reuses cached objects to render the scene. In the Random walk moving pattern, the direction changes randomly so, the probability to visit the same object is small.

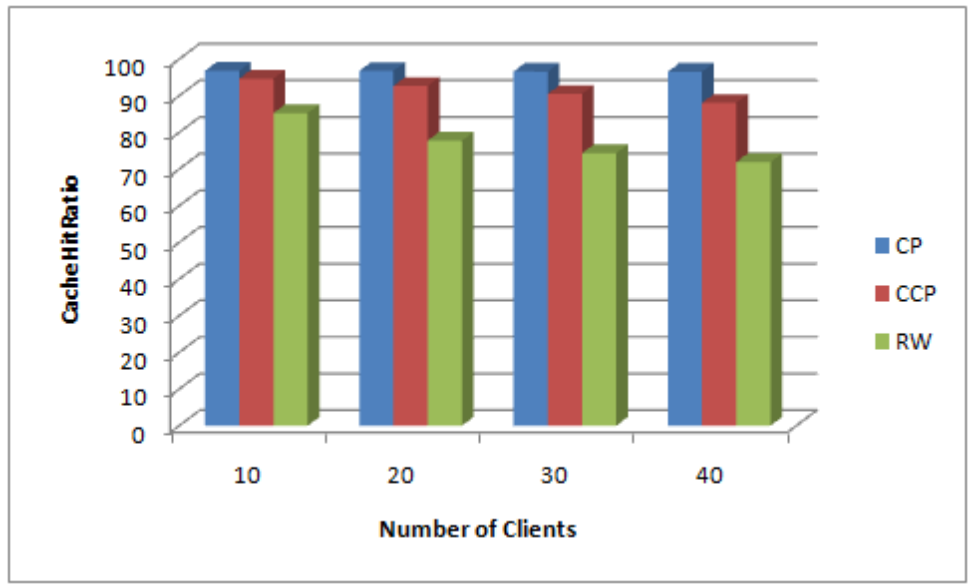

Figure 33: Average cache hit ratio for CP, CCP and RW in type II mobiles

For type II mobiles, the movement pattern has insignificant effect on the cache hit ratio because the client always requests base mesh only, which always exist in cache. Therefore, the cache can store a huge number of objects. We can conclude that, the guided navigation seems to give better performance than the random ones on mobiles with high resolution.

### 4.2.5 Experiment 5

This experiment evaluates the efficiency of the proposed framework through combining the three main modules together; progressive mesh, caching technique and constrained object's resolution request. The experiment settings are explained in table 8.

| Parameters | Values |
|---|---|
| **Number of clients** | 10, 20, 30,40 clients |
| **Cache size** | 2MB |
| **Type of mobile** | Using type I & type II |

Table 8: Experiment 5 Parameters

As shown in table 9, the average response time using type I of mobiles increases gradually with the raises of the number of clients from 10 until 40 clients. This is because, as the number of clients increases, the number of requests grows and hence, the wireless medium becomes busier.

Table 9: Simulation performance using Type I mobiles

| Users<br>performance measure | 10 | 20 | 30 | 40 |
|---|---|---|---|---|
| Average response time(s) | 0.0817 | 0.26659 | 0.334 | 0.403 |
| Average no. of requests per minute | 132.9 | 684.2 | 818.3 | 919.3 |
| Average wireless medium utilization % | 11.24 | 25.31 | 34.62 | 41.89 |
| Average Cache hit Ratio % | 84.9 | 77.6 | 74.3 | 73.6 |

Type II of mobile achieves more improvement than type I as shown in table 10. This is due to the fact that type II of clients almost requests the base mesh only, which leads to significantly reducing the amount of requests and therefore positively affecting the response time, user perception, network traffic, wireless utilization and cache hit ratio.

Table 10 : Simulation performance using Type II mobiles

| Users<br>Performance measures | 10 | 20 | 30 | 40 |
|---|---|---|---|---|
| Average response time(s) | 0.01820 | 0.1827 | 0.20446 | 0.30133 |
| Average no. of requests per minute | 42.5 | 51.4 | 66.2 | 91.2 |
| Average wireless medium utilization % | 0.082 | 0.232 | 2.942 | 5.502 |
| Average Cache hit Ratio % | 96.78 | 96.75 | 96.58 | 96.51 |

## 4.2.6 Real World Experiments

### 4.2.6.1 Experiment 1

Now, we will discuss a case study that has been designed to proof the concept of our simulation mentioned above. The case study has been implemented on real power phone. Table 10 contains the real experiment setting.

Table 11: Real Experiment Parameters

| Parameters | Values |
|---|---|
| **Number of clients** | One client |
| **Number of Dynamic objects** | 10 3D objects |
| **Type of mobile** | Galaxy tab model GT-P1000 (type II) |
| **CPU** | with CPU 1 GHz Catex-A8 |
| **RAM** | RAM 512 |
| **Cache size** | 2 MB from the total cache size |
| **OS** | Android 2.2 (Froyo ) |
| **Tools** | Opengl ES 1.1 API and java 1.7 |
| **Type of network** | Open network |

The Circular movement pattern is used. The movement pattern is pre-processed and used in the framework as input. The scene consists of ten 3d objects represented by progressive mesh and some static objects. Figure 34 shows a scene snapshot.

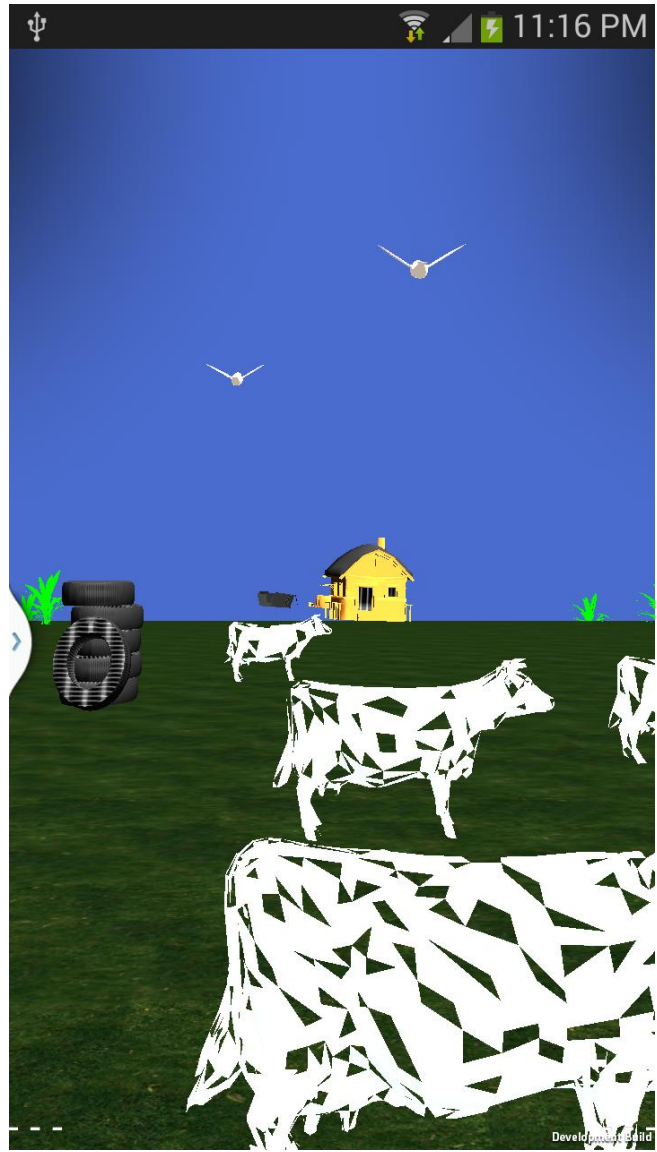

**Figure 34: Screenshot from Real world Experiment**

The case study has the same scenario of our simulation. We used the case study to measure the same metrics measured by the simulation such as response and latency time; cache hit ratio, virtual perception, number of requests per minutes and number of frame per second.

Figures 35 and 36 depict the average response time and latency time after running the mobile application for 8 minutes.

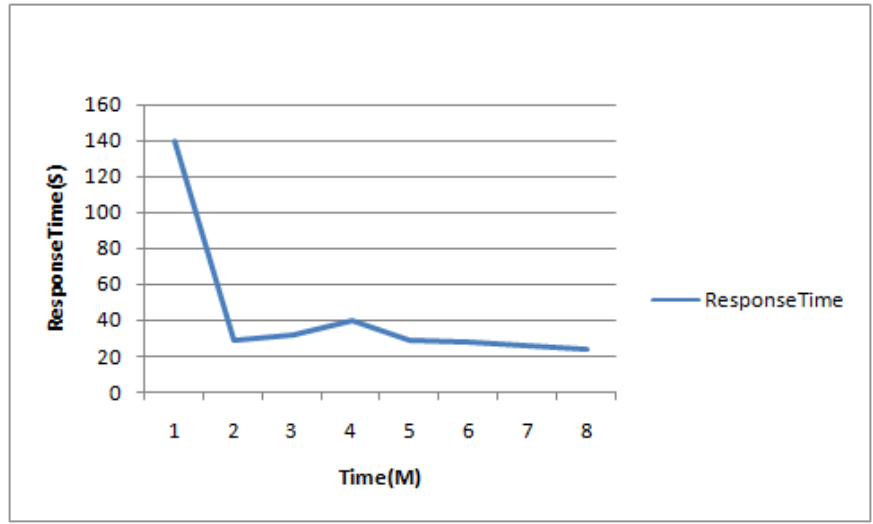

**Figure 35: Average Response Time**

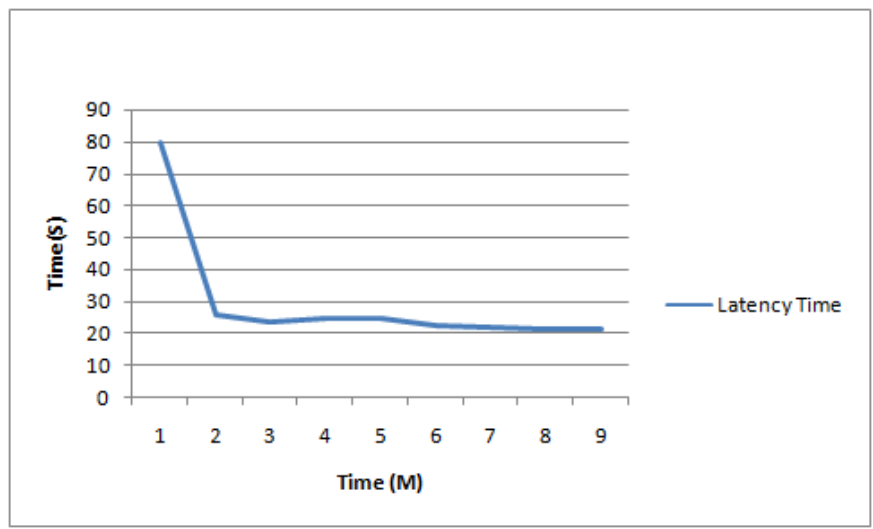

**Figure 36: Average Latency Time**

As shown in figures 35 and 36, the first minute indicates high value of response time because it includes the initialization time to connect to the server. The time before minute 6 changes according to the cache hit ratio and the number of demand requests, after that the time starts to get stable because the cache includes a large number of resolutions for each object, and then the number of requests decreases as a result. As depicted in figures 37 and 38, the cache hit ratio percentage is increased gradually until it reaches 100% then the cache starts to remove some object's resolution cached in it to allocate free space for the new incoming objects.

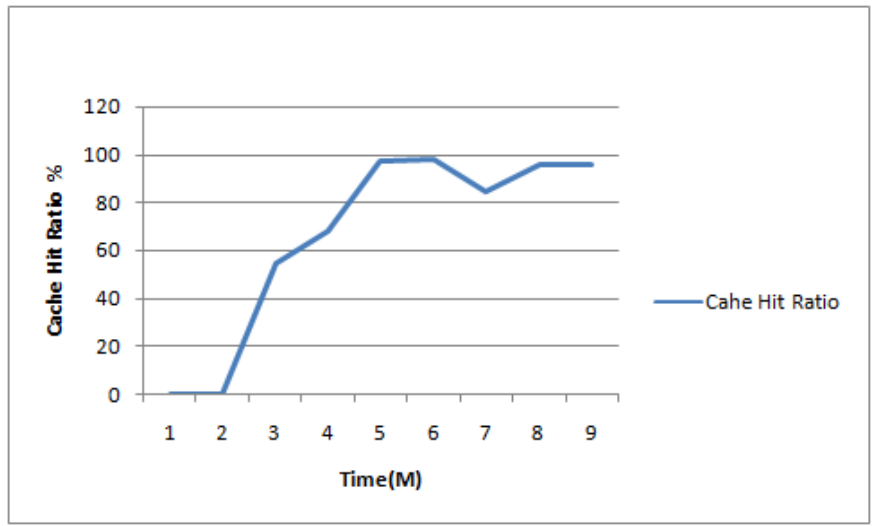

Figure 37:   Cache Hit Ratio

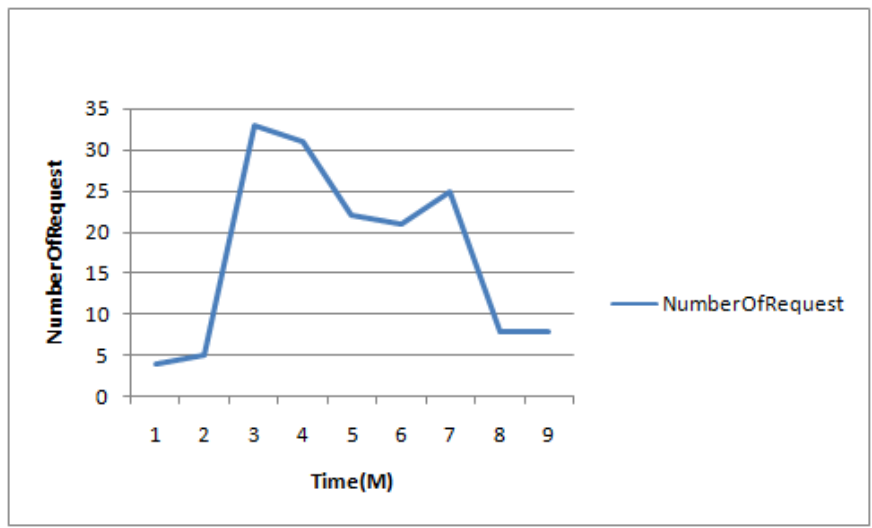

Figure 38:   Number of Requests per Minute

The above discussion is valid for virtual perception metric that depends on the ratio of cached objects' resolutions to the optimal object's resolution as shown in figure 39.

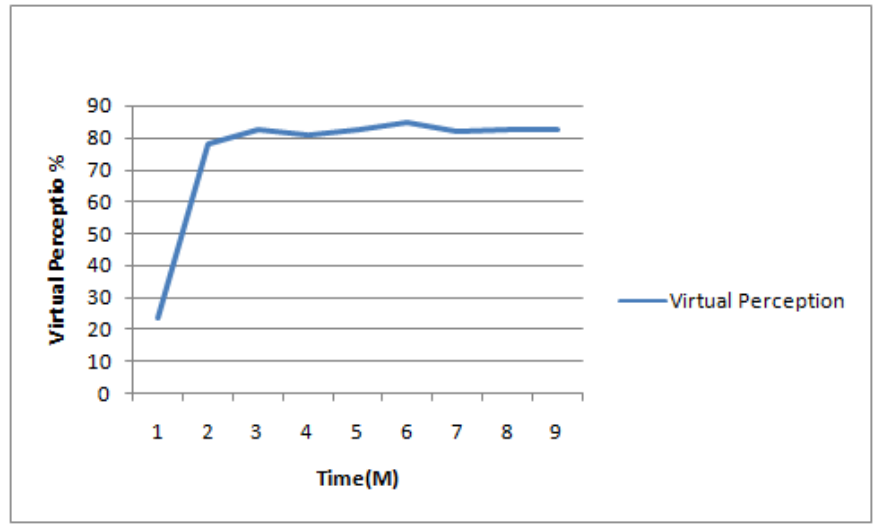

Figure 39: Virtual Perception

A frame per second (fps) is one of the most important metrics for smooth navigation through the virtual environment. It depends on the number of points rendered on the mobile's device display, therefore, objects of 10% resolution has smaller number of points to be rendered than these of 20, 30,... until 80% resolution. It is shown in figure 40, that fps decreases as object's resolution increases.

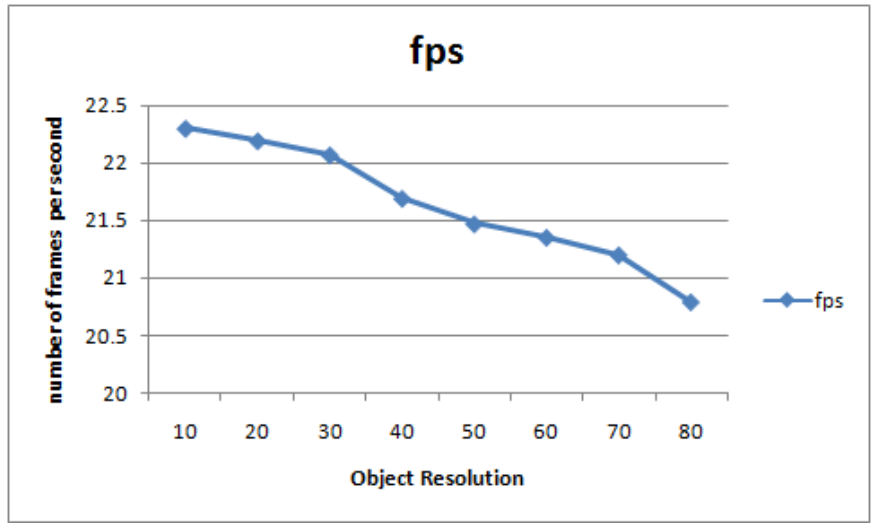

Figure 40: Frame per Second FPS

### 4.2.6.2 Experiment 2 (Power consumption Experiment)

This experiment evaluates the effect of using progressive mesh on mobile power consumption. We used a smart phone that is known for its short battery life time. Table 12 shows the experiment setting.

Table 12: Power consumption Experiment Setting

| Parameters | Values |
|---|---|
| **Number of client** | 1 client |
| **Number of Dynamic 3D objects** | 15 3D objects |
| **Type of mobile** | Samsung galaxy sIII |
| **CPU** | Quad. Core 1.4 GHz cortex. Ag |
| **RAM** | 1 GB |
| **Battery** | li- long 2 100 mAh battery Take Time : up to 21 h 40 Min (2G)/ up to 11 h 40 min (3G) |
| **OS** | Android Kit Kat |
| **Software Tools** | Battery Monitor Widget version 3.1.7.1 |

This experiment measures the display and transmission effects using progressive mesh on power consumption through the mobile device.

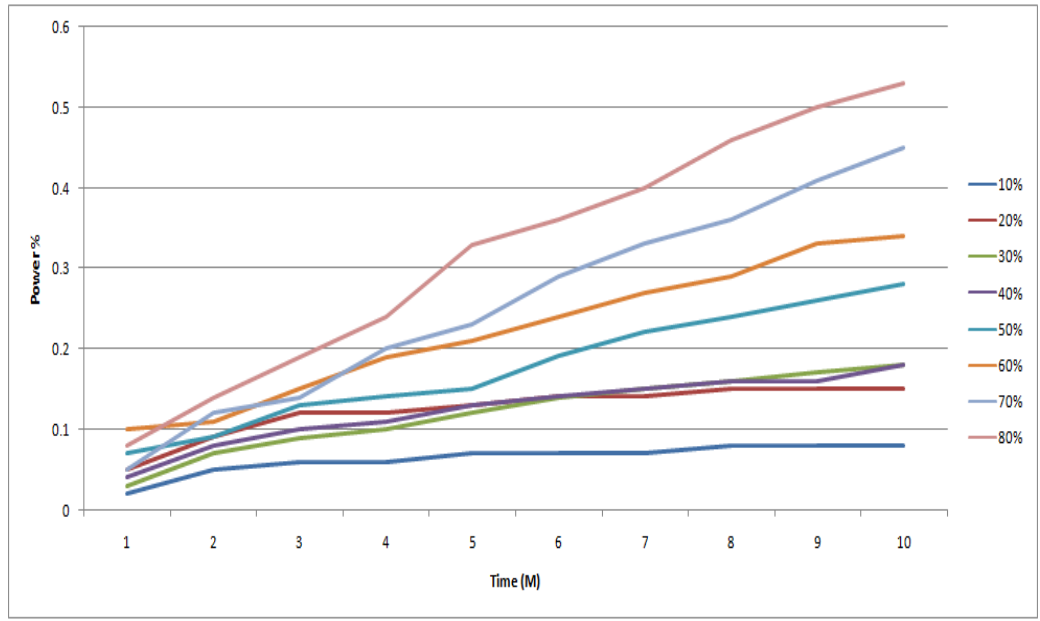

**Figure 41: Power Consumption**

As shown in figure 41 the battery consumption increases with the increase of 3D object resolutions. In 10 % object resolution, the line is nearly stable from the first minutes until to the last minute but at 80% object resolution it has a linear representation. Figure 41 proves that using progressive mesh leads to decreasing the power consumption. In the static level of details each time the client sends a request for small resolution the server replies with the nearest resolution to the requested one. Rendering and transmission of a high resolution each time leads to consuming the mobile battery.

## 4.3 Simulation and Real world

A real mobile application has been developed to evaluate the performance of using progressive mesh technique, caching technique and constrained object's resolution through a walkthrough mobile application in the proposed framework. We measured response time, latency time, cache hit ratio, virtual perception, number of requests per minute and frames per second. Depicted in the results, the simulation allowed us to test different mobile configurations and different settings to expect what will happen in a real world. Figures 42 and 43 show the results we obtained from both the real mobile application and the simulation, as taken from figures 19, 23 from experiment1 and figures 35, 38 from the real experiment. As predicted in figure 42, the response time in both graphs has the same trend.

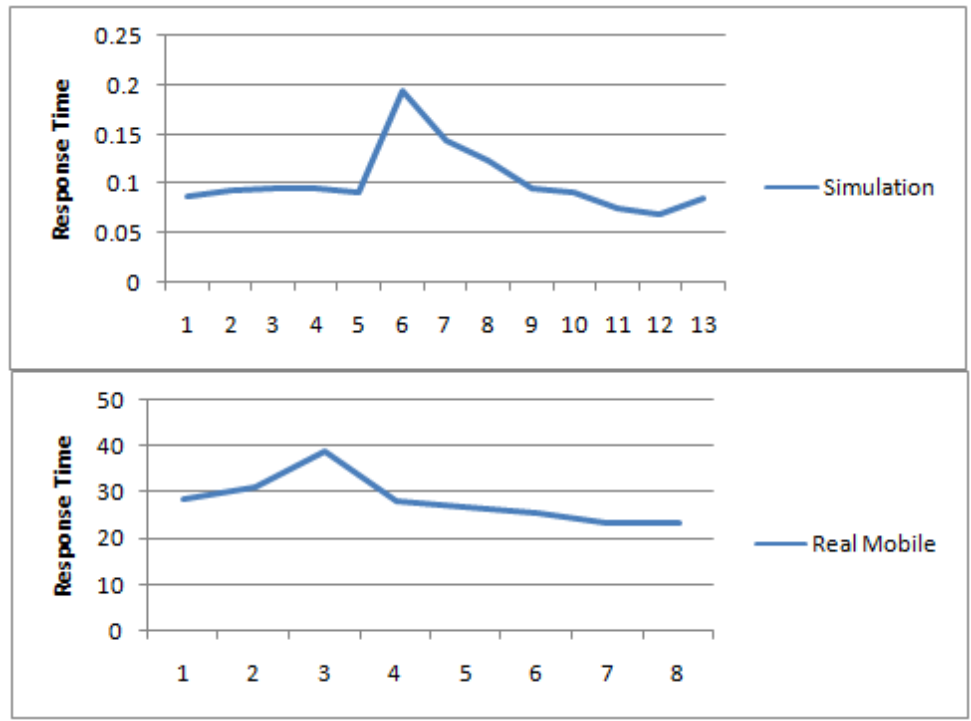

Figure 42: Response Time in simulation and real mobile

In both the simulation and real experiments, the response time starts stable then, when the density of requests increases in the middle of run time it has a peak, and after that, it decreases. The same trend is applied to the average number of requests shown in figure 43 which proves the goal of the simulation.

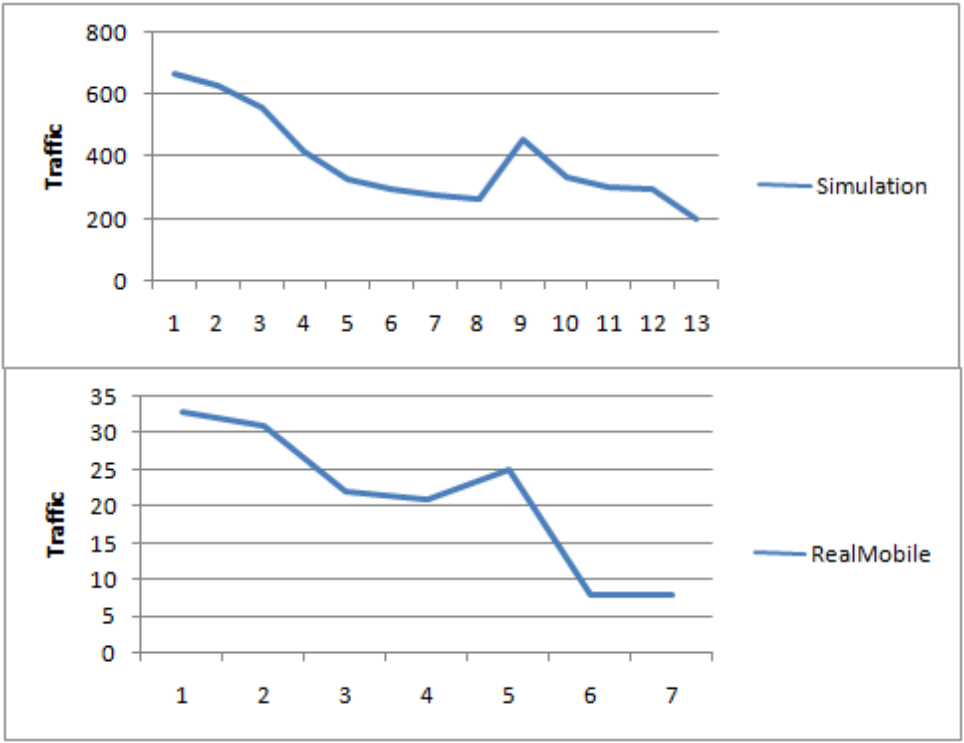

**Figure 43: traffic in Simulation and Real Mobile**

# CHAPTER 5
# CONCLUSION AND FUTURE WORK

This chapter presents our conclusions and ideas for further research. Section 5.1 summarizes our framework and its effect on the mobile performance. Section 5.2 focuses on several optimizations to improve the proposed framework.

# 5 Conclusions and Future Work

## 5.1 Conclusion

This thesis presents an efficient framework for running interactive applications on mobiles phones in wireless medium. We developed a framework for an optimized mobile walkthrough application by integrating progressive mesh, caching, and constrained object's request techniques.

In the previous chapters we showed how we used those techniques to improve our framework to overcome the limitation of the phone's power consumption and the network, namely:

- Using the progressive mesh technique to minimize data transfer through the wireless medium and to improve the real time interaction. The thesis included a comparison between progressive mesh techniques and using static level of details.

- Further optimization is achieved by caching previously fetched objects to reduce the amount of requests sent by the user.

- Finally, we constrain the client's requests according to device specifications. Hence, the device requests only the actual number of progressive records that are needed to render and display the scene. Consequently, it improves the utilization of both mobile battery and network traffic.

The experiments that have been developed focused evaluating the following:

- Computing progressive mesh overhead in transmission and rendering (Response time, wireless utilization, traffic, frame per second),

- Measuring the speed of interaction on mobile screen (virtual perception, latency time),

- Reckon the cache hit ratio and its effects on mobile framework

- Score the power consumption

Our results showed that using progressive mesh improved 64% of the average system response time, 70% of user perception, 80% of latency and 91% of cache hit ratio compared to its static level of details counterpart. Using constrained object's resolution request improved 33% of network traffic, 57% of Wireless medium utilization and 87% of response time. Finally using cache improved 90% of response time, 35% of traffic and 68% of wireless medium utilization.

We developed a case study using a real mobile phone to prove the simulation results and we had the same trend between the simulation results and the real world experiment using a mobile device.

## 5.2 Future work

In our future work, we are planning to extend the framework to include the following:

- ✓ Implementing all simulation experiments on the real mobiles to further prove the simulation concepts.
- ✓ Running the framework on different mobiles using different mobile specification.
- ✓ Studying the impact of our framework on the power consumption and extending it with energy saving techniques.
- ✓ Gathering feedback from real users to evaluate the visual perception metric.
- ✓ And finally we will use cloud computing approaches to enhance the proposed framework performance

# تقنية الواقع الافتراضى للهواتف الجوالة خلال الاطار اللا سيلكى

إعداد

غادة محمد فتحى محمد

**تحت اشراف**

أ.د. ريم بهجت

أ.د. ولاء شتا

رسالة مقدمة إلي كلية الحاسبات والمعلومات

جامعة القاهرة

كجزء من متطلبات الحصول علي

درجةالماجستير في علوم الحاسب

كلية الحاسبات والمعلومات

جامعة القاهرة

جمهورية مصر العربية

**2014**

# إقرار

أقر بأن هذا العمل لم يسبق قبوله لأي درجة علمية وليس مقدما حاليا للحصول علي أي درجة علمية أخري .

كما أنه تم إسناد كل ما هو مقتبس في هذه الرسالة إلي مصدرها لأصلي مع ذكر المصادر بوضوح تام.

**اسم الطالب :** غادة محمد فتحى

**التوقيع:**

# ملخص الرسالة

قد شهدت السنوات القليلة الماضية نموا كبيرا فى عدد متنوع من الواقع الافتراضى على الهواتف النقالة والتى تسمح للمستخدمين التفاعل والتنقل خلال المواقع التاريخية والمتاحف والمدن الافتراضية من خلال هواتفهم النقالة .

وتدعم هذه التطبيقات العديد من العملاء وتفرض شروطا ثقيلة على موارد الشبكة اللاسلكية والموارد الحاسوبيةِ.

ومن احد الاسباب الرئيسية فى تصميم تطبيقات فعالة للهواتف النقالة هى نقل البيانات بين الخادم والاجهزة النقالة.

هذا البحث يطرح إطار فعال لأنتقال وتخزين الشبكة التقدمية والتى تخزن وتقسم الكائنات ثلاثية الابعاد الى قرارات مختلفة . فى هذا النهج كل جهاز نقال يتلقى تدريجيا الكائنات الخاصة به و التى تتماشى خصائصها مع امكانيات ومواصفات الجهاز النقال وبذلك يتحسن وقت الاستجابة النظام العام والادراك للمستخدم.

ويطرح الإطار نظام لتخزين الكائنات ثلاثية الابعاد و الذى يستخدم للحفاظ على التفاصيل الأكثر طلبا فى ذاكرة الهاتف النقال مما يؤدي الى تقليل حركة المرور داخل الشبكة اللاسلكية .

ولقد تمت التجارب على بيئة المحاكاة و العالم الحقيقى لتثبت فاعلية الاطار المقترح على الاعدادات مختلفة للبيئة الافتراضية وعلى اجهزة نقالة متنوعة .

أظهرت النتائج التجريبية ان الأطار المقترح تمكن من تحسين اداء الواقع الافتراضى على الهواتف النقالة مع وجود قيود صغيرة نسبيا .

# نبذه عن فصول الرساله

نظمت الرسالة على النحو التالى:

- يعرض الفصل الاول فكرة عامة عن الاطار المقترح , كما انه يوضح الدوافع للبدء فى العمل على هذا الاطار كما انه يوفر دليل للقارىئ.

- يوفر الفصل الثانى خلفية عن الاعمال ذات صلة لجميع البحوث التى تمت فى مجال الهواتف النقالة كما انه يعطى خلفية عامة عن جميع الطرق المستخدمة فى بناء الاطار .

- الفصل الثالث يوضح مع الشرح المفصل كيف حققنا الاهداف من تصميم الاطار كما انه يشرح تصميم الاطار فى كل من الجانبين الخادم والهاتف .

- يظهر الفصل الرابع الدراسات التجريبية التى اجريت لتقيم الاطار المقترح ويقدم شرح مساهمات الرسالة فى مجال الهواتف النقالة .

- يلخص الفصل الخامس مساهمات الرسالة واستنتاجات العمل كما انه يسلط الضوء على بعض الاتجاهات المستقبلية التى يمكن تحقيقها .